\documentclass[11pt]{article}
\usepackage[T1]{fontenc}
\usepackage[numbers,square]{natbib}
\usepackage[utf8]{inputenc}
\usepackage{booktabs}
\IfFileExists{lmodern.sty}{\usepackage{lmodern}}{}
\usepackage{amsmath,amssymb,amsthm,mathtools,bm}
\usepackage{enumitem,microtype,xcolor}
\usepackage{comment}
\usepackage[letterpaper,margin=1in]{geometry}
\usepackage{abstract}

\usepackage[unicode,colorlinks=true,linktoc=all,linkcolor=blue!40!black,citecolor=blue!40!black,urlcolor=blue!40!black]{hyperref}
\IfFileExists{algorithm.sty}{\usepackage{algorithm}\usepackage{algpseudocode}}{%
  \usepackage{float}\newfloat{algorithm}{tbp}{loa}\floatname{algorithm}{Algorithm}%
  \newenvironment{algorithmic}{\begin{list}{}{\leftmargin=2em\itemsep=0pt\parsep=0pt}}{\end{list}}%
}
\providecommand{\State}{\item}
\providecommand{\Require}{\item[\textbf{Input:}]}

\providecommand{\Return}{\textbf{return } }
\providecommand{\While}[1]{\item[\textbf{while }#1\textbf{ do}]}
\providecommand{\EndWhile}{\item[\textbf{end while}]}
\providecommand{\If}[1]{\item[\textbf{if }#1\textbf{ then}]}
\providecommand{\Else}{\item[\textbf{else}]}
\providecommand{\EndIf}{\item[\textbf{end if}]}
\allowdisplaybreaks[1]
\numberwithin{equation}{section}
\newtheorem{theorem}{Theorem}[section]
\newtheorem{lemma}[theorem]{Lemma}
\newtheorem{proposition}[theorem]{Proposition}
\newtheorem{corollary}[theorem]{Corollary}
\theoremstyle{remark}\newtheorem{remark}[theorem]{Remark}
\newcommand{\vct}[1]{\boldsymbol{#1}}

\newcommand{\ten}[1]{\boldsymbol{\mathsf{#1}}}
\newcommand{\dd}{\mathrm d}
\newcommand{\sstar}{s_{\star}}
\newcommand{\fstar}{f_{\star}}
\DeclareMathOperator{\Reg}{Reg}
\title{Breaking the One-Player Extension Barrier: History-Aware Repair for Multiplayer Approximate Nash Equilibria}

\author{
	Jinqi Tang\thanks{
		School of Mathematical Sciences,
		Peking University.
		Email: \texttt{tangjinqi74@gmail.com}.
	}
	\qquad
	Shihong Ding\thanks{
		School of Intelligence Science and Technology,
		Peking University.
		Email: \texttt{dingshihong@stu.pku.edu.cn}.
	}
	\qquad
	Cong Fang\thanks{
		School of Intelligence Science and Technology,
		Peking University.
		Email: \texttt{fangcong@pku.edu.cn}.
	}
}

\date{}

\begin{document}

\maketitle

\begin{abstract}
Computation of Nash equilibria is a central problem in
algorithmic game theory, yet exact equilibrium computation is 
$\mathsf{PPAD}$-complete even for two-player games.
This 
hardness motivates the
study of polynomial-time algorithms for approximate
Nash equilibria.
While substantial progress has been made for
two-player and three-player games,
existing algorithms for general $m$-player games have largely relied on repeatedly applying one-player extension starting from improved fixed-player algorithms. 
Such improvements can at best strengthen the classical $1-1/m$ guarantee to $1-1/(m-c)$, for some constant $c$ independent of $m$. We break this long-standing barrier by developing an iterative algorithm that directly tackles general $m$-player games and achieves a $(1-(\sigma_{\star}-\epsilon)/m)$-approximation for fixed $\epsilon > 0$ and sufficiently large $m$, where $\sigma_\star\approx1.5867$
is the unique root in $(1,2)$ of
$\sigma_\star^3-19\sigma_\star^2+27\sigma_\star+1=0$.
Our approach rests on a novel iterative descent-and-repair framework that allows multiple rounds of repair. At the heart of our analysis is a new potential function inspired by continuous-time dynamics, which certifies progress across iterations and ensures polynomial-time termination. Our results highlight the power of history-aware, iterative repair in breaking approximation barriers for multiplayer games.
\end{abstract}

\clearpage 


\section{Introduction}
\label{sec:intro}
Nash equilibrium is a fundamental concept in game theory.
Although every finite game admits a Nash equilibrium~\cite{Nash1951}, computing an exact equilibrium is
computationally challenging.
In particular, the problem is $\mathsf{PPAD}$-complete
even for two-player games~\cite{ChenDeng2006,ChenDengTeng2009,DGP2009},
and $\mathsf{FIXP}$-complete for games with three
or more players~\cite{EtessamiYannakakis2010}.
These complexity results motivate the study of efficient
algorithms for computing approximate Nash equilibria. For games with payoffs normalized to $[0,1]$, an $\varepsilon$-approximate Nash equilibrium is a mixed-strategy profile in which no player can improve her expected payoff by more than $\varepsilon$ through a unilateral deviation. A central question is how small an approximation error can be guaranteed in polynomial time. 

For bimatrix games, a long sequence of results has progressively improved polynomial-time approximation guarantees through different algorithmic techniques. Early approaches relied primarily on direct strategy constructions and small-support mixing, achieving approximation guarantees of $3/4$~\cite{KontogiannisEtAl2009} and $1/2$~\cite{DMP2006}. Daskalakis, Mehta, and Papadimitriou~\cite{DMP2007} subsequently moved beyond constant-size supports, using sampling and linear programming to obtain $(3-\sqrt{5})/2$. Another approach exploited auxiliary zero-sum games: Bosse, Byrka, and Markakis~\cite{BBM2010} used an equilibrium of a suitably constructed zero-sum game as a starting point and refined the resulting strategies through linear programming, achieving $1/2-1/(3\sqrt{6})$.

A major conceptual advance was introduced by Tsaknakis and Spirakis~\cite{TS2008}, who formulated approximate Nash equilibrium computation as an optimization problem over the maximum regret function. Their framework uses a descent procedure to find an approximate stationary point, followed by a final strategy construction, yielding a $0.3393$-approximation. This paradigm of descending to an approximate stationary point remained influential, with subsequent refinements improving the final mixed-strategy adjustment while still performing it only once. Deligkas, Fasoulakis, and Markakis~\cite{DFM2022} enriched the candidate strategies used in the final construction, improving the guarantee to $1/3$. Subsequent developments in automated approximation analysis and optimal strategy mixing~\cite{DengLiLi2025Automation,DengLiLi2025OptimalMixing} have further strengthened this line of research, leading to recent guarantees of approximately $0.30954$ and $0.30919$~\cite{LiLi2026,Gondauri2026}. These later improvements largely retain the stationary-point-based approach, obtaining stronger guarantees through more sophisticated strategy constructions and mixing procedures.

However, extending the optimization-based framework of Tsaknakis and Spirakis~\cite{TS2008} to general multiplayer games presents significant challenges. A major reason is that a single mixed-strategy adjustment often fails to yield substantial improvements in the multiplayer setting. Approximate stationarity provides only first-order information, whereas simultaneously mixing multiple players' strategies introduces higher-order interaction terms in the multi-affine payoff expansion. As the number of players grows, these interactions become increasingly complex. For three-player games, Li, Li, and Deng~\cite{LiLiDeng2026LegoNE} recently extended the stationary-point-based approach through \emph{LegoNE}, an automated framework that, together with a reasoning large language model, discovered a polynomial-time $(0.5+\delta)$-approximation algorithm. Notably, even for three-player games, the analysis of its worst-case approximation guarantee heavily relies on computer-assisted optimization, highlighting the difficulty of analyzing such constructions.

A natural and standard approach to general multiplayer games is to start with an algorithm for a fixed number of players and repeatedly apply the one-player extension technique developed by Briest et al.~\cite{BGR2008} and H\'emon et al.~\cite{HRS2008}. This technique transforms an $\alpha$-approximation for $r$-player games into a $1/(2-\alpha)$-approximation for $(r+1)$-player games (Lemma~\ref{lem:recursive-extension}), yielding the classical $1-1/m$ approximation guarantee for general $m$-player games (Lemma~\ref{lem:init}). Even when starting from substantially improved fixed-player algorithms, repeated extension can at best yield a guarantee of the form $
1-1/(m-c),
$
for some constant $c$ independent of $m$ (Proposition~\ref{prop:recursive-limitation}). Consequently, such improvements affect only lower-order terms, leaving the leading coefficient of $1/m$ unchanged.
Other works have investigated approximate Nash equilibria in multiplayer games using different ideas. Czumaj, Fasoulakis, and Jurdzi\'nski~\cite{CFJ2017} studied relative (multiplicative) approximation guarantees, while small-support and sampling-based approximations have also been explored~\cite{HRS2008,BBP2014,BBP2017}. Arieli and Babichenko~\cite{ArieliBabichenko2017} investigated connections between approximate equilibria and discrepancy theory. Further results have been obtained for structured game classes, including anonymous, polymatrix, and aggregative games~\cite{DaskalakisPapadimitriou2015,GoldbergTurchetta2017,DeligkasEtAl2017Polymatrix,Babichenko2018Aggregative}.

More recently, Deng, Li, Wang, and Zhang~\cite{DengLiWangZhang2026Hardness} show that, under deterministic PPAD-ETH, for every fixed $\varepsilon\in(0,1)$ there exists a constant player count $p_\varepsilon$ for which computing an $\varepsilon$-approximate Nash equilibrium requires quasipolynomial worst-case time in the complete payoff table's encoding length. Their result rules out a polynomial-time approximation guarantee with a fixed error for multiplayer games under the same hypothesis. But this leaves open how much the classical $1-1/m$ guarantee can be improved as the number of players grows. Since one-player extension from a fixed-player algorithm leaves the leading coefficient of $1/m$ unchanged, this raises a central question: 

\begin{quote}
\emph{
Is there a polynomial-time algorithm
for general $m$-player games that goes beyond one-player
extension and improves the leading coefficient of
the $1-1/m$ approximation guarantee?
}
\end{quote}

Theorem~\ref{thm:main} answers this question affirmatively.
Specifically, our proposed algorithm improves the leading
coefficient of $1/m$ from $1$ to any fixed value
below $\sigma_\star \approx 1.5867$.
This improves on the $1-1/(m-c)$ guarantee.

\subsection{Our Result}
Recent algorithmic advances in approximate Nash equilibria for a small, fixed number of players increasingly rely on computer-assisted analyses, offering limited insight into the general multiplayer setting. In contrast, we develop a polynomial-time
algorithm, called \textbf{I}terative \textbf{D}escent and \textbf{R}epair with \textbf{H}istory
(\textbf{IDRH}), that directly tackles general $m$-player games
and achieves an asymptotic approximation guarantee strictly
better than that of one-player extension. This iterative algorithm exploits historical information to enable multiple rounds of repair, supported by a self-contained analytic proof. 

\begin{theorem}[Approximate Nash Equilibria in Multiplayer Games]
\label{thm:main}
Let $\sigma_\star$ be the unique root in $(1,2)$ of
\[
    \sigma_\star^3
    -19\sigma_\star^2
    +27\sigma_\star+1=0,
    \qquad
    \sigma_\star\approx 1.5867.
\]
For fixed rational \(\sigma\in(1,\sigma_\star)\), there exists a computable threshold \(m_0(\sigma)\) such that, for every \(m\ge m_0(\sigma)\), our algorithm computes a rational \((1-\sigma/m)\)-approximate Nash equilibrium of an \(m\)-player game with rational payoffs in \([0,1]\), in time polynomial in the input bit length.
\end{theorem}
We assume rational payoffs, as is standard in computational game theory~\cite{ChenDengTeng2009,EtessamiYannakakis2010}, enabling exact arithmetic and polynomial bit-complexity analysis. Many real-world strategic interactions involve large numbers of players, motivating the study of approximate equilibria in such games~\cite{BBP2014,ArieliBabichenko2017,GempEtAl2022}. Our result is asymptotic in $m$, with the cutoff $m_0(\sigma)$ in~\eqref{eq:finite-player-cutoff} accounting for finite-size errors and depending only on $\sigma$.

\subsection{Overview of Our Techniques}

We next outline the ideas behind Algorithm~\ref{alg:main}. Each iteration begins with a linearized descent LP step inspired by the optimization procedure of Tsaknakis and Spirakis~\cite{TS2008}. However, just a stationary profile with only one final adjustment need not meet the target approximation guarantee. Instead, we propose alternating \emph{descent} and \emph{repair} throughout the algorithm (Subsections~\ref{subsec:algorithm-descent}, \ref{subsec:algorithm-repair}). The challenge is to retain useful information across repairs and ensure global progress even when the maximum regret is not monotone. We construct a novel \emph{potential function}~\eqref{eq:analysis-finite-potential} motivated by the continuous-time dynamics of the algorithm (Corollaries~\ref{cor:analysis-descent-ode}, \ref{cor:analysis-repair-ode}). The following three paragraphs sketch our techniques.

\paragraph{Linearized Descent LP.}
In each step, the algorithm begins with the linearized descent LP in Subsection~\ref{subsec:algorithm-descent}, inspired by Tsaknakis and Spirakis~\cite{TS2008}, which searches for a simultaneous strategy update that decreases the maximum regret. Nonnegative weights $z_{j,b}$ with $\sum_{j,b}z_{j,b}\le1$ define a descent direction, applied with a small step size $\eta$. When a sufficiently improving direction exists, Subsection~\ref{subsec:analysis-dual} guarantees quantitative progress, with higher-order interaction terms controlled through the multi-affine structure of the payoffs. Otherwise, LP duality yields the certificate~\eqref{eq:algorithm-dual-law}: a distribution over players and associated mixed deviations. The resulting replacement inequalities~\eqref{eq:analysis-dual-columns} constrain strategy updates and provide the fresh information used in repair.

\paragraph{Dual--History Reweighted Repair.}
To build on this dual information, the repair in Subsection~\ref{subsec:algorithm-repair} combines fresh certificates with retained historical information. Previous stationary-point-based algorithms use dual information primarily for a final strategy construction~\cite{TS2008,DFM2022,LiLiDeng2026LegoNE}. Here we retain the contributions of earlier repairs and use them to guide later activations. The reweighting rule~\eqref{eq:repair-controls} assigns less new activation, relative to the current dual weights, to players with more retained historical weight. After best-response adjustments and corrections, the history update retains old contributions and records the new ones. This reuse of history makes successive repairs work together toward the target guarantee (Subsection~\ref{subsec:analysis-history}).

\paragraph{A Potential Function.}
The final challenge is to demonstrate that the algorithm makes progress in both the descent and repair branches, thereby guaranteeing termination in polynomial time. To this end, we introduce a novel potential function to measure joint progress: descent reduces regret but discounts history, whereas repair creates history without necessarily reducing regret. Inspired by the continuous-time perspective of Su, Boyd, and Cand\`es~\cite{SuBoydCandes2016} on Nesterov's accelerated gradient method, we study the infinitesimal behavior of scaled satisfaction $\varsigma=m(1-\Reg(\vct{x}))$ and historical weight $H$. This leads to the initial form of the potential function. Let $\Phi(\varsigma,H)=\varsigma/H+F(H)$. The successful-descent drift determines the shape of $F$, ensuring that the loss of historical weight is balanced by the gain in satisfaction. The repair drift then determines the achievable coefficient, yielding $\sigma_\star\approx1.5867$. We discuss the details in Subsection~\ref{subsec:analysis-potential} and further introduce the finite-step corrections in Subsection~\ref{subsec:analysis-finite}.

\paragraph{Organization of the Paper.}
Section~\ref{sec:pre} introduces the basic notation, reviews the classical one-player extension technique, and establishes its asymptotic limitation. Section~\ref{sec:algo} presents our iterative descent-and-repair algorithm, including the linearized descent LP and dual-history reweighted repair mechanism. Section~\ref{sec:analysis} gives a proof sketch of the main theorem, explaining repeated repairs, the potential function, and the improved coefficient $\sigma_\star$. Section~\ref{sec:detailed-proofs} provides the complete technical proofs, including finite-step error bounds, parameter verification, and polynomial bit-complexity analysis. Section~\ref{sec:conclusion} concludes the paper.


\section{Problem Setup and Preliminaries}
\label{sec:pre}
In this section, we fix the game model and notation, then review
one-player extension as a starting point for our algorithm.
We consider a normal-form game with player set $[m]=\{1,\ldots,m\}$.
Each player $i$ has a nonempty finite action set $A_i$ and an explicitly
listed payoff tensor
$\ten{U}_i\in([0,1]\cap\mathbb{Q})^{A_1\times\cdots\times A_m}$.
The binary encoding length of the entire input is denoted by $B$.

A mixed profile is $\vct x=(\vct x_1,\ldots,\vct x_m)$, where
$\vct x_i\in\Delta(A_i)$ and
$\Delta(A_i)$ denotes the set of all probability
distributions over the pure action set $A_i$. We use $u_i(\vct x)$ to represent the corresponding
multilinear expected payoff, i.e., $u_i(\vct x) = \langle \ten{U}_i,\vct x_1\otimes \vct x_2\otimes \cdots \otimes \vct x_m\rangle $. For a mixed action $\vct y_j$, write
$\vct x[j\leftarrow\vct y_j]:=(\vct y_j,\vct x_{-j})$ for the profile obtained by replacing
$\vct x_j$ by $\vct y_j$. We identify a pure action $a\in A_j$ with its
unit vector $\vct e_a$. The regret of player $i$ and the maximum regret
are, respectively,
$$ 
  r_i(\vct x)
  :=\max_{a\in A_i}u_i(\vct x[i\leftarrow\vct e_a])-u_i(\vct x),
  \qquad
  \Reg(\vct x):=\max_i r_i(\vct x).
$$ 
Thus $\vct x$ is an $\varepsilon$-approximate Nash equilibrium precisely
when $\Reg(\vct x)\le\varepsilon$.
For $i\in[m]$ and $a\in A_i$, define
$$ 
  g_{i,a}(\vct x)
   :=u_i(\vct x[i\leftarrow\vct e_a])-u_i(\vct x),
  \qquad
  d_{i,a}(\vct x):=1-g_{i,a}(\vct x).
$$ 
Because payoffs lie in $[0,1]$, every $d_{i,a}(\vct x)$ is nonnegative.
We extend $d_{i,a}(\vct x)$ affinely to a mixed deviation
$\vct b_i\in\Delta(A_i)$, writing
$ d_{i,\vct b_i}(\vct x)=
 1-u_i(\vct x[i\leftarrow\vct b_i])+u_i(\vct x)$.
Set
$$ 
  q_i(\vct x):=\min_{a\in A_i}d_{i,a}(\vct x)=1-r_i(\vct x),
  \qquad
  q(\vct x):=\min_iq_i(\vct x)=1-\Reg(\vct x).
$$ 
For Algorithm~\ref{alg:main}, we use superscripts to denote iteration indices. At iteration $k$, we write
\begin{equation}\label{eq:pre-scaled-satisfaction}
  \sigma^{(k)}:=mq(\vct x^{(k)})
     =m\bigl[1-\Reg(\vct x^{(k)})\bigr].
\end{equation}
The prescribed target is denoted by $\sigma$, whereas
$\sigma^{(k)}$ is the value attained by the current profile.
The stopping condition $\sigma^{(k)}\ge\sigma$ is therefore
 equivalent to $\Reg(\vct x^{(k)})\le1-\sigma/m$.

We next recall the classical one-player extension technique for
approximate Nash equilibria in Lemma~\ref{lem:recursive-extension}.

\begin{lemma}[One-Player Extension~\cite{HRS2008}]
\label{lem:recursive-extension}
Suppose that a rational $\alpha$-approximate $(\alpha \in \mathbb{Q}\cap [0,1])$ Nash
equilibrium can be computed in polynomial time
for $r$-player games. 
Then a rational $1/(2-\alpha)$-approximate Nash
equilibrium can be computed in polynomial time
for $(r+1)$-player games.
\end{lemma}

\begin{proof}
Fix a pure action $s$ of player $r+1$ and compute a rational
$\alpha$-approximate equilibrium $\vct x$ of the $r$-player game
obtained by fixing that action. Let $b$ be a pure best response of
player $r+1$ to $\vct x$, and put $\theta=1/(2-\alpha)$.
Output the first $r$ marginals $\vct x$ together with
$\theta\vct e_s+(1-\theta)\vct e_b$ for the new player.
For each original player the regret is at most
$\theta\alpha+(1-\theta)=\theta$: against $s$ it is bounded
by $\alpha$ and against $b$ by one. The new player's regret is
at most $\theta$, since $b$ is her best response. 
\end{proof}

Repeatedly applying Lemma~\ref{lem:recursive-extension} yields the
classical $1-1/m$ approximation guarantee. As recorded in
Lemma~\ref{lem:init}, this gives the initial profile for Algorithm~\ref{alg:main}.

\begin{lemma}[Rational Initialization]
\label{lem:init}
For every rational
$m$-player game with payoffs in $[0,1]$,
a rational product profile $\vct x^{(0)}$
satisfying $\sigma^{(0)}\ge 1$
can be computed in polynomial time.
\end{lemma}

\begin{proof}
For $m=1$, a pure best response gives an exact
Nash equilibrium.
For $m\ge 2$, 
repeated application of Lemma~\ref{lem:recursive-extension}
gives approximation factors satisfying
$
    \alpha_{r+1}=\frac{1}{2-\alpha_r}.
$
By induction, $\alpha_r=1-1/r$.
Hence the resulting rational profile satisfies
$$ 
    \Reg(\vct x^{(0)})\le 1-\frac1m,
    \qquad
    \sigma^{(0)}\ge 1.
$$
The construction is polynomial in the explicitly
listed input size, and all mixing probabilities
are rational.
\end{proof}

The scope of this construction is clarified by
Proposition~\ref{prop:recursive-limitation}, which shows that improving the algorithm for any fixed number of players leaves the asymptotic leading coefficient of recursive extension unchanged. This motivates the sufficiently-large-$m$ regime of Theorem~\ref{thm:main}, where we improve that coefficient by directly addressing the multiplayer setting.

\begin{proposition}[Limitation of Recursive Extension]
\label{prop:recursive-limitation}
Fix an integer $r$ and an approximation
guarantee $0\le\alpha_r<1$ for $r$-player games.
Repeated application of the one-player extension
rule yields, for every $m\ge r$, an approximation
guarantee of
\[
    \alpha_m=1-\frac{1}{m-c},
    \qquad
    c:=r-\frac{1}{1-\alpha_r}.
\]
Thus, recursive extension cannot improve the
leading coefficient of $1/m$, even when starting
from an improved approximation guarantee for
a fixed number of players.
\end{proposition}

\begin{proof}
By the one-player extension rule in Lemma~\ref{lem:recursive-extension},
$
    \alpha_{k+1}=\frac{1}{2-\alpha_k}.
$
Rearranging gives
$
    \frac{1}{1-\alpha_{k+1}}
    =\frac{1}{1-\alpha_k}+1.
$
Iterating from $r$ to $m$, we obtain
\[
    \frac{1}{1-\alpha_m}
    =\frac{1}{1-\alpha_r}+m-r
    =m-c.
\]
Hence $\alpha_m=1-1/(m-c)$.
\end{proof}


\section{The IDRH Algorithm}
\label{sec:algo}
In this section, we describe the iterative updates of Algorithm~\ref{alg:main}.
The algorithm maintains a product profile $\vct x^{(k)}$
and historical weights $\vct h^{(k)}$ retained from earlier repairs.
At each iteration, it solves the linearized descent LP in Subsection~\ref{subsec:algorithm-descent}
and uses the outcome to choose the next update: success gives the
descent update, while failure triggers
the history-aware repair of Subsection~\ref{subsec:algorithm-repair}.
We clarify the choice of some important parameters in our algorithm in Subsection~\ref{subsec:algorithm-parameters}, and more detailed discussions are deferred to Subsection~\ref{subsec:proof-controls}. 

\begin{algorithm}[t]
\caption{\textbf{I}terative \textbf{D}escent and
         \textbf{R}epair with \textbf{H}istory (\textbf{IDRH})}
\label{alg:main}
\begin{algorithmic}[1]
\Require An explicit rational game and a fixed rational target
$1<\sigma<\sigma_\star$ described in Section~\ref{sec:pre}. 
\State Choose $d,\eta,u$ by Subsection~\ref{subsec:algorithm-parameters} and the
       remaining controls and cutoff $m_0$ in Subsection~\ref{subsec:proof-controls}.
\State Construct $\vct x^{(0)}$ by Lemma~\ref{lem:init} and compute
       $\sigma^{(0)}$ by~\eqref{eq:pre-scaled-satisfaction}.
\State Initialize $k\gets 0$, $\vct h^{(0)}\gets\vct 0$, and
       $H^{(0)}\gets 0$.
\If{$m<m_0$}
    \State \Return $\vct x^{(0)}$.
\EndIf
\While{$\sigma^{(k)}<\sigma$}
    \State Compute $\mu^{(k)}$ by~\eqref{eq:algorithm-threshold}
           in Subsection~\ref{subsec:algorithm-parameters} and solve the linearized descent LP~\eqref{eq:linearized-lp}.
    \If{$z_{\star} \le \Reg(\vct x^{(k)})-\eta\mu^{(k)}/m$} \Comment{Descent branch}
        \State Compute $P^{(k)}$ and update the profile
               by~\eqref{eq:algorithm-descent-update}.
        \State Set $\vct h^{(k+1)}\gets P^{(k)}\vct h^{(k)}$ and
               $H^{(k+1)}\gets P^{(k)}H^{(k)}$.
    \Else 
        \Comment{Repair branch}
        \State Extract $\vct\lambda^{(k)}$ and
               $(\vct b_i^{(k)})_i$ by~\eqref{eq:algorithm-dual-law}.
        \State Compute the activation weights and reverse budgets
               by~\eqref{eq:algorithm-history}
               and~\eqref{eq:repair-controls}.
        \State Activate the deviations by~\eqref{eq:algorithm-activation}
               and perform the reverse sweep~\eqref{eq:algorithm-reverse-sweep}.
        \State Freeze the deficits by~\eqref{eq:algorithm-correction}
               and perform the forward correction~\eqref{eq:algorithm-forward-correction}.
        \State Update $\vct h^{(k+1)}$ and $H^{(k+1)}$
               by~\eqref{eq:algorithm-repair-memory}.
    \EndIf
    \State Compute $\sigma^{(k+1)}$ by~\eqref{eq:pre-scaled-satisfaction} and increment $k$.
\EndWhile
\State \Return $\vct x^{(k)}$.
\end{algorithmic}
\end{algorithm}

\subsection{Linearized Descent LP}
\label{subsec:algorithm-descent}

We first formulate the descent LP~\eqref{eq:linearized-lp}, which searches
for a simultaneous strategy update that reduces the maximum regret.
At the current profile $\vct x^{(k)}$, let $\eta>0$ be the fixed
rational step size specified in~\eqref{eq:algorithm-primary-controls}
and let $\mu^{(k)}>0$ be the margin defined
in~\eqref{eq:algorithm-threshold} of Subsection~\ref{subsec:algorithm-parameters}. For each player $j$ and action $b$,
write
\[
    \Delta_{j,b}g_{i,a}(\vct x)
    :=g_{i,a}(\vct x[j\leftarrow\vct e_b])-g_{i,a}(\vct x).
\]
This is the change in player $i$'s gain from deviating to action $a$
when player $j$'s strategy is replaced by the pure action $b$.
We solve
\begin{equation}\label{eq:linearized-lp}
\begin{aligned}
 \min_{\vct z,z_\star}\quad & z_\star\\
 \text{subject to}\quad
 & z_{j,b}\ge 0,\qquad \sum_{j,b}z_{j,b}\le 1,\\
 & g_{i,a}(\vct x^{(k)})
   +\eta\sum_{j,b}z_{j,b}\Delta_{j,b}g_{i,a}(\vct x^{(k)})
   \le z_\star
   \qquad \text{for all }i,a.
\end{aligned}
\end{equation}
The linearized descent LP~\eqref{eq:linearized-lp} is motivated by the descent method of Tsaknakis and Spirakis~\citep{TS2008}. 
We seek a feasible direction that decreases the maximum regret
$\Reg(\vct{x})=\max_{i,a}g_{i,a}(\vct{x})$. At iteration $k$, any
nonnegative weights $\vct{z}$ satisfying $\sum_{j,b}z_{j,b}\le 1$
define a direction $\vct{p}(\vct{z})=(\vct{p}_j(\vct{z}))_{j=1}^m$,
where $\vct{p}_j(\vct{z})=\sum_b z_{j,b}(\vct{e}_b-\vct{x}_j^{(k)})$.
Since each deviation gain $g_{i,a}$ is affine in each player's strategy,
its directional derivative at $\vct{x}^{(k)}$ along $\vct{p}(\vct{z})$
is $\sum_{j,b}z_{j,b}\Delta_{j,b}g_{i,a}(\vct{x}^{(k)})$.
Adding $\eta$ times this derivative to the
current gain gives the linearized gain in~\eqref{eq:linearized-lp}.

LP~\eqref{eq:linearized-lp} \emph{succeeds} if its optimum is at most
$\Reg(\vct x^{(k)})-\eta\mu^{(k)}/m$, where $\mu^{(k)}$ is defined
in~\eqref{eq:algorithm-threshold}. In this case, take the optimal
rational solution $\vct z^{(k)}$ and set
\begin{equation}\label{eq:algorithm-descent-update}
\begin{gathered}
 Z_j^{(k)}:=\sum_b z_{j,b}^{(k)},\qquad
 Z^{(k)}:=\sum_j Z_j^{(k)},\qquad
 P^{(k)}:=\prod_j(1-\eta Z_j^{(k)}),\\
 \vct x_j^{(k+1)}
 :=(1-\eta Z_j^{(k)})\vct x_j^{(k)}
   +\eta\sum_b z_{j,b}^{(k)}\vct e_b.
\end{gathered}
\end{equation}
In the product mixture~\eqref{eq:algorithm-descent-update}, coordinate $j$ draws from its old strategy
with probability $1-\eta Z_j^{(k)}$. Thus $P^{(k)}$ is the
probability that every coordinate draws from its old strategy.
We discount the historical weights by this factor:
\begin{equation}\label{eq:algorithm-descent-memory}
    h_i^{(k+1)}:=P^{(k)}h_i^{(k)}
    \quad\text{for all }i,\qquad
    H^{(k+1)}:=P^{(k)}H^{(k)}.
\end{equation}

If LP~\eqref{eq:linearized-lp} \emph{fails}, we use the dual
certificates~\eqref{eq:algorithm-dual-law} to select deviations for
repair in Subsection~\ref{subsec:algorithm-repair}, following the use of dual solutions to construct candidate
strategies in bimatrix games~\citep{TS2008,DFM2022}.
Let $w_{i,a}\ge 0$ be the weights on the deviation constraints in an
optimal rational dual solution, with $\sum_{i,a}w_{i,a}=1$.
By complementary slackness, if $w_{i,a}>0$, then deviation $(i,a)$
attains the largest linearized gain at the optimal LP direction.
In particular, $a$ is a best response to the mixture of opponents'
profiles used in the linearization. This mixture gives total weight
at most $\eta$
to single-coordinate replacements, so the expected
payoff of any pure action differs by at most $\eta$ from its payoff
against the current opponents' strategies $\vct{x}_{-i}^{(k)}$.
Thus, every action with $w_{i,a}>0$ is a $2\eta$-best response to
$\vct{x}_{-i}^{(k)}$. For repair, we mix these approximate best
responses in proportion to their weights in the dual certificate.
Specifically, we set
\begin{equation}\label{eq:algorithm-dual-law}
 \lambda_i^{(k)}:=\sum_a w_{i,a},\qquad
 \vct{b}_i^{(k)}
 :=\frac{\sum_a w_{i,a}\vct{e}_a}{\lambda_i^{(k)}}
 \quad\text{when }\lambda_i^{(k)}>0.
\end{equation}
When $\lambda_i^{(k)}=0$, we take $\vct{b}_i^{(k)}$ to be the
smallest-indexed pure action. Subsection~\ref{subsec:algorithm-repair}
uses these mixed deviations together with historical information to
repair the current strategy profile.

\subsection{Dual-History Reweighted Repair}
\label{subsec:algorithm-repair}

Next, we use the dual information from a failed descent step to construct
a repair. At iteration $k$, the dual law~\eqref{eq:algorithm-dual-law}
supplies a player distribution $\vct\lambda^{(k)}$ and mixed deviations
$\vct b_i^{(k)}$. We then combine this information with $\vct h^{(k)}$
through the reweighting rule~\eqref{eq:repair-controls}, which determines
the activation weights and reverse budgets.

\paragraph{Historical Information and Activation Weights.}
We first specify how retained history influences the new activation weights.
The historical weight $h_i^{(k)}$ records how much of the weight
assigned to player $i$'s deviations in earlier repairs is retained
in the current profile.
Let $\lambda_i^{(k)}$ be the weight assigned to player $i$ by the
current dual certificates. We define the total historical weight,
each player's centered weight, and the average of the centered weights
under the current player distribution by
\begin{equation}\label{eq:algorithm-history}
 H^{(k)}:=\sum_i h_i^{(k)},\qquad
 v_i^{(k)}:=m h_i^{(k)}-H^{(k)},\qquad
 \tau^{(k)}:=\sum_i\lambda_i^{(k)}v_i^{(k)}.
\end{equation}
Thus $v_i^{(k)}$ is $m$ times the difference between player $i$'s
historical weight and the mean $H^{(k)}/m$. When $H^{(k)}>0$, we
write $\psi^{(k)}:=\tau^{(k)}/H^{(k)}$ for the difference between
the $\vct\lambda^{(k)}$-weighted mean of the historical weights and
the uniform mean, expressed as a fraction of $H^{(k)}/m$.

The satisfaction bound~\eqref{eq:rawrepair} combines new activations,
retained activations, and a reverse best-response sweep. Let $\beta$ and $\nu$ be the fixed
coefficients for new and retained activations, respectively, in the
lower bound on each player's satisfaction. Let $\rho$ be the
probability that no player takes the best-response branch in the
reverse sweep, so the total budget for that sweep is $1-\rho$.
The exact choices of $\rho$, $\beta,\nu$, and the reweighting
constants $G_0,Q$ are given in~\eqref{eq:finite-retention-controls},
\eqref{eq:finite-repair-coefficients}, and~\eqref{eq:finite-reweighting-controls},
respectively, in Subsection~\ref{subsec:proof-controls}.

Let $\gamma_i^{(k)}$ denote player $i$'s weight for the new activation,
$D^{(k)}$ the total contribution before interaction losses, and
$\delta_i^{(k)}$ player $i$'s share of the reverse-sweep budget.
Their definitions are
\begin{equation}\label{eq:repair-controls}
\begin{aligned}
 \gamma_i^{(k)}
 &:=\lambda_i^{(k)}
    \frac{G_0-Qv_i^{(k)}}{G_0-Q\tau^{(k)}},\\
 D^{(k)}
 &:=(1-\rho)+\beta+\nu H^{(k)}
   =\beta(G_0+QH^{(k)}),\\
 \delta_i^{(k)}
 &:=\frac{D^{(k)}}{m}-\beta\gamma_i^{(k)}-\nu h_i^{(k)}.
\end{aligned}
\end{equation}
In the activation rule~\eqref{eq:repair-controls}, the factor $G_0-Qv_i^{(k)}$ decreases as
player $i$'s retained weight increases. Its average under
$\vct\lambda^{(k)}$ is $G_0-Q\tau^{(k)}$, which gives the
normalizing denominator. Thus, among players with equal positive
dual weights, the one with more historical weight receives less
new activation weight. In particular, $\gamma_i^{(k)}<\lambda_i^{(k)}$
exactly when $\lambda_i^{(k)}>0$ and $v_i^{(k)}>\tau^{(k)}$.
If all historical weights are zero, $\vct\gamma^{(k)}=\vct\lambda^{(k)}$.
Lemma~\ref{lem:finite-legal-tilt} shows that $\vct\gamma^{(k)}$ is
a probability distribution.

To see how the budget is allocated, consider the two activation
terms in the satisfaction bound of Lemma~\ref{lem:rawrepair}.
The term $\beta\gamma_i^{(k)}$ counts the current activation when
zero or one opponent takes a new branch. For a retained activation,
the case in which all players keep their base strategies is already
accounted for in the old profile. The additional term $\nu h_i^{(k)}$
comes from cases in which one opponent takes a new branch.
Summing the two terms over players gives $\beta+\nu H^{(k)}$.
The reverse sweep supplies the remaining $1-\rho$, giving the total
$D^{(k)}$. Its allocation $\delta_i^{(k)}$ fills the gap between
player $i$'s two activation terms and the common share $D^{(k)}/m$.
Lemma~\ref{lem:finite-legal-tilt} also ensures that these allocations
are nonnegative and sum to $1-\rho$.

\paragraph{Activation and Reverse Best-Response Sweep.}
Next, we apply the activation weights and reverse budgets to construct
the raw repair profile. With activation intensity $t$ from~\eqref{eq:finite-retention-controls},
player $i$ uses her deviation with probability $t\gamma_i^{(k)}$, giving
the strategy
\begin{equation}\label{eq:algorithm-activation}
 \vct y_i^{(k)}
  :=(1-t\gamma_i^{(k)})\vct x_i^{(k)}
    +t\gamma_i^{(k)}\vct b_i^{(k)}.
\end{equation}
For the reverse sweep, use the budgets in~\eqref{eq:repair-controls} to set
\begin{equation}\label{eq:algorithm-reverse-sweep}
 P_0^{(k)}:=1,\qquad
 P_i^{(k)}:=1-\sum_{j\le i}\delta_j^{(k)},\qquad
 p_i^{(k)}:=\frac{\delta_i^{(k)}}{P_{i-1}^{(k)}}.
\end{equation}
Process the players in
the order $m,m-1,\ldots,1$. At player $i$, compute a pure best
response to $\vct y_j^{(k)}$ for $j<i$ and to the already updated
$\widetilde{\vct x}_j^{(k)}$ for $j>i$. Form $\widetilde{\vct x}_i^{(k)}$ by
mixing this best response into $\vct y_i^{(k)}$ with probability
$p_i^{(k)}$.

To see why~\eqref{eq:algorithm-reverse-sweep} gives player $i$ her assigned
share, observe that $1-p_j^{(k)}=P_j^{(k)}/P_{j-1}^{(k)}$. Thus the probability
that every player $j<i$ keeps $\vct y_j^{(k)}$ is
$\prod_{j<i}(1-p_j^{(k)})=P_{i-1}^{(k)}$. The event that these players keep
$\vct y_j^{(k)}$ and player $i$ uses her reverse best response has
probability $P_{i-1}^{(k)}p_i^{(k)}=\delta_i^{(k)}$. Players $j>i$ are unrestricted
on this event, so the conditional distribution of the opponents'
actions is exactly the one used to compute the best response.

\paragraph{Forward Correction and Retention of Historical Information.}
Finally, we correct the remaining shortfalls in satisfaction and update
the historical weights.
In the raw profile $\widetilde{\vct x}^{(k)}$ produced by
\eqref{eq:algorithm-activation} and~\eqref{eq:algorithm-reverse-sweep}, all players keep their
pre-sweep strategies $\vct y_i^{(k)}$ with probability
$\prod_i(1-p_i^{(k)})=P_m^{(k)}=\rho$. Conditional on this event, they all use
their original strategies $\vct x_i^{(k)}$ with probability
$\prod_i(1-t\gamma_i^{(k)})$, by independence across coordinates.
Multiplying gives $C_{\rm raw}^{(k)}$, the probability that every coordinate
uses its original branch through both stages, as recorded
in~\eqref{eq:algorithm-correction}.

On this event the original profile contributes at least
$C_{\rm raw}^{(k)}\sigma^{(k)}/m$ to each player's satisfaction $q_i(\widetilde{\vct x}^{(k)})$.
Adding the common share $D^{(k)}/m$ gives the target for the raw repair.
Before changing any strategy further, measure each player's shortfall
from this target as follows.
\begin{equation}\label{eq:algorithm-correction}
 C_{\rm raw}^{(k)}:=\rho\prod_i(1-t\gamma_i^{(k)}),\qquad 
 e_i^{(k)}
 :=\left(\frac{C_{\rm raw}^{(k)}\sigma^{(k)}+D^{(k)}}{m}
       -q_i(\widetilde{\vct x}^{(k)})\right)_+,\qquad
 E^{(k)}:=\sum_i e_i^{(k)}.
\end{equation}
Keep the shortfalls in~\eqref{eq:algorithm-correction} fixed and process the players in the order
$1,2,\ldots,m$. At player $i$, mix a pure best response to the
current opponents into her strategy with probability
\begin{equation}\label{eq:algorithm-forward-correction}
 \theta_i^{(k)}:=\frac{e_i^{(k)}}{1+\sum_{j\le i}e_j^{(k)}}.
\end{equation}
This gives the corrected profile $\vct x^{(k+1)}$.

By~\eqref{eq:algorithm-forward-correction},
$1-\theta_i^{(k)}=(1+\sum_{j<i}e_j^{(k)})/(1+\sum_{j\le i}e_j^{(k)})$,
so the probability that no player takes a forward correction is
$\prod_i(1-\theta_i^{(k)})=1/(1+E^{(k)})$. The probability of retaining all
original branches through the entire repair is therefore
$\widehat C^{(k)}=C_{\rm raw}^{(k)}/(1+E^{(k)})$. The controls guarantee
$\widehat C^{(k)}\ge c$ (Lemma~\ref{lem:finite-raw-error}).
An old record can thus be retained with factor $c$. For the new
record of player $i$, replacing only her base activation label by her
deviation label changes the event probability to
$\widehat C^{(k)}\,t\gamma_i^{(k)}/(1-t\gamma_i^{(k)})\ge c\,t\gamma_i^{(k)}$. We therefore
record $c\,t\gamma_i^{(k)}$ for this repair and multiply the old weight
by $c$. With $c$ and $t$ as in~\eqref{eq:finite-retention-controls},
the update is
\begin{equation}\label{eq:algorithm-repair-memory}
 h_i^{(k+1)}:=c\bigl(h_i^{(k)}+t\gamma_i^{(k)}\bigr)
 \quad\text{for all }i,\qquad
 H^{(k+1)}:=c\bigl(H^{(k)}+t\bigr),
\end{equation}
where the last equality uses $\sum_i\gamma_i^{(k)}=1$.

\subsection{Fixed Control Parameters}
\label{subsec:algorithm-parameters}

To complete the algorithm description, we specify the three main controls
for a fixed rational target $1<\sigma<\sigma_\star$: the reserve $d$
in the descent margin~\eqref{eq:algorithm-threshold},
the descent step size $\eta$ in~\eqref{eq:algorithm-descent-update},
and the repair step scale $u$ underlying~\eqref{eq:finite-retention-controls}. Choose
\begin{equation}\label{eq:algorithm-primary-controls}
 d\in\mathbb Q_{>0},\qquad \sigma+8d<\sigma_\star,\qquad 
 \eta=\frac12\min\left\{\frac14,\frac d{24}\right\},\qquad
 u=\frac12\min\left\{\frac1{144},\frac d{100}\right\}.
\end{equation}
The reserve accommodates finite-step errors, while $\eta$ and $u$
control the sizes of the two types of update. All three depend only
on $\sigma$. Subsection~\ref{subsec:proof-controls} gives an effective
choice of $d$ and defines the remaining fixed controls and the
player-count cutoff $m_0$.

The descent margin depends on the current profile and the total
historical weight through
\begin{equation}\label{eq:algorithm-threshold}
 S_u(H):=1+H+R_0\frac{1+H}{1+QH},\qquad
 \mu^{(k)}:=S_u(H^{(k)})-\sigma^{(k)}-d.
\end{equation}
The ratios $R_0$ and $Q$ are defined in~\eqref{eq:finite-reweighting-controls}.


\section{Analysis of the Algorithm}
\label{sec:analysis}
In this section, we give a proof sketch of Theorem~\ref{thm:main}.
We first track how the two update branches change scaled satisfaction
$\varsigma$ and retained history $H$ in Subsections~\ref{subsec:analysis-dual}, \ref{subsec:analysis-history}. These estimates guide the construction
of the potential function in Subsection~\ref{subsec:analysis-potential},
which we then use to establish progress and termination. Further finite-step corrections and running time analysis are discussed in Subsection~\ref{subsec:analysis-finite}. Subsection~\ref{subsec:main-proof} organizes the preceding arguments. 

For the one-step analysis, fix a nonterminal iteration $k$ of
Algorithm~\ref{alg:main} with $m\ge m_0(\sigma)$. Write
$\vct{x}=\vct{x}^{(k)}$, $\vct{h}=\vct{h}^{(k)}$, and
$H=\sum_i h_i$ for the current profile, the vector of historical
weights, and its total weight, respectively. We use
$\varsigma=\sigma^{(k)}=m(1-\Reg(\vct{x}))$ for the current scaled
satisfaction and reserve $\sigma$ for its target value. 

\subsection{Successful Descent LP}
\label{subsec:analysis-dual}

We begin with the successful-LP branch, where the strategy update
improves satisfaction while discounting retained history.
Proposition~\ref{prop:descent} bounds these one-step changes, and passing
to the limit as $\eta\to0$ gives the infinitesimal drifts in
Corollary~\ref{cor:analysis-descent-ode}.

\begin{proposition}[Finite-Step Guarantee for a Successful LP]
\label{prop:descent}
If the descent LP~\eqref{eq:linearized-lp} succeeds, let $Z_j=\sum_b z_{j,b}$,
$Z=\sum_j Z_j\le 1$, and $P=\prod_j(1-\eta Z_j)$ for the selected
solution $\vct{z}$. Then $\vct{h}^+=P\vct{h}$, $H^+=PH$, and
\begin{equation}\label{eq:main-descent}
 \varsigma^+\ge P\bigl[(1+\eta Z)\varsigma+\eta\mu\bigr]
 \ge \varsigma+\eta\mu-\eta^2(\varsigma+\mu).
\end{equation}
\end{proposition}

\begin{proof}
Fix a deviation $(i,a)$ and write $d_0=d_{i,a}(\vct{x})$ and
$d_{j,b}=d_{i,a}(\vct{x}[j\leftarrow\vct{e}_b])$. In the product
update~\eqref{eq:algorithm-descent-update}, the event that all coordinates use their base strategies has
probability $P$. The event that only coordinate $j$ draws a non-base
label, specifically $b$, has probability $P\eta z_{j,b}/(1-\eta Z_j)$.
Every conditional
deficit is nonnegative, so these events give
$d_{i,a}(\vct{x}^+)\ge P(d_0+\eta\sum_{j,b}z_{j,b}d_{j,b})$.
The successful test for LP~\eqref{eq:linearized-lp} says that
$(1-\eta Z)d_0+\eta\sum_{j,b}z_{j,b}d_{j,b}\ge(\varsigma+\eta\mu)/m$.
Adding $\eta Z d_0\ge\eta Z\varsigma/m$ and taking the minimum over
deviations gives the first inequality in~\eqref{eq:main-descent}.
The second follows from $P\ge1-\eta Z$, $Z\le1$, and $\mu>0$.
The all-base event retains every contribution to $\vct{h}$ with
factor $P$.
\end{proof}

\begin{corollary}[ODEs for Successful LP Steps]
\label{cor:analysis-descent-ode}
Fix the current state and the controls that determine $\mu$
in~\eqref{eq:algorithm-threshold}. Along
successful steps with $\eta\downarrow0$, pass to a subsequence on
which $Z\to z\in[0,1]$. On the time scale of one step of length
$\eta$, the scalar differential relations are
\begin{equation}\label{eq:analysis-descent-ode}
 \dot H=-zH,\qquad \dot\varsigma\ge S_u(H)-\varsigma-d .
\end{equation}
Here and below, $\dot H$ denotes the limit of the one-step difference
quotient for $H$, and $\dot\varsigma$ its lower-limit counterpart
for $\varsigma$.
\end{corollary}

\begin{proof}
Since $\sum_j Z_j\le1$, the product satisfies
$P=1-\eta Z+O(\eta^2)$ uniformly in the LP solution. Substituting into
$H^+=PH$ gives $(H^+-H)/\eta\to-zH$. Dividing the second bound
in~\eqref{eq:main-descent} by $\eta$ after subtracting $\varsigma$
gives $\liminf_{\eta\downarrow0}(\varsigma^+-\varsigma)/\eta\ge\mu$.
\end{proof}

\subsection{Repair Step When the LP Fails}
\label{subsec:analysis-history}

Next, we analyze the repair step after a failed LP. To use this branch
in the progress argument, we need both feasibility and bounds on the
resulting changes in satisfaction and history. We establish these in
Proposition~\ref{prop:repair}, then derive the first-order drifts as
$u\to0$ in Corollary~\ref{cor:analysis-repair-ode}. Detailed proofs are given in Subsection~\ref{subsec:failed-repair}. 

\begin{proposition}[Finite-Step Guarantee for Dual-History Reweighted Repair]
\label{prop:repair}
At a nonterminal iteration, a failed descent LP~\eqref{eq:linearized-lp} yields a
feasible repair from its dual certificates and the historical information. Let $E$ be the total
correction deficit in~\eqref{eq:algorithm-correction} and put
$D=\beta(G_0+QH)$ as in~\eqref{eq:repair-controls}. Then
\begin{equation}\label{eq:main-repair}
 \varsigma^+\ge c\varsigma+D-E,\qquad
 0\le E\le\frac{120u}{m},\qquad
 \vct{h}^+=c(\vct{h}+t\vct{\gamma}),\qquad H^+=c(H+t).
\end{equation}
The update preserves $H^+\le L_0$ and
$mh_i^+-H^+\le R_0H^+$ for each player $i$.
\end{proposition}

\begin{proof}
For each earlier repair $s<k$, let $A^{(s,k)}$ denote its common
retained mass. It starts at $A^{(s,s+1)}=c$ and is multiplied at
each later iteration $\ell$ by $P^{(\ell)}$ on descent or by $c$ on
repair. Lemma~\ref{lem:certificate-coupling} shows that the retained
activation events are disjoint and preserve their original conditional
action laws.

By Lemma~\ref{lem:descent}, the failed LP yields dual certificates
such that, for every $j$ and $\vct{y}_j\in\Delta(A_j)$,
\begin{equation}\label{eq:analysis-dual-columns}
 F_j(\vct{y}_j):=\sum_i\lambda_i d_{i,\vct{b}_i}
   (\vct{x}[j\leftarrow\vct{y}_j])
 <\frac{\varsigma+\mu}{m},
 \qquad \lambda_j\le F_j(\vct{b}_j).
\end{equation}
Combining~\eqref{eq:analysis-dual-columns} with the historical overlap estimate of
Lemma~\ref{lem:history-overlap} gives the feasibility
condition~\eqref{eq:finite-floor-feasibility}.
Lemma~\ref{lem:finite-legal-tilt} then guarantees that
$\vct{\gamma}$ is a probability vector, the reverse allocations are
nonnegative and sum to $1-\rho$, and the repair preserves both
historical weight bounds.

Lemma~\ref{lem:finite-raw-error} gives $0\le E\le120u/m$ and
$C_{\rm raw}/(1+E)\ge c$. Since $0\le D\le2u<1$, the corrected
bound~\eqref{eq:corrected-bound} yields
$\varsigma^+\ge c\varsigma+D/(1+E)\ge c\varsigma+D-E$.
The retention bound in Lemma~\ref{lem:finite-raw-error} permits the exact updates
$\vct{h}^+=c(\vct{h}+t\vct{\gamma})$ and $H^+=c(H+t)$
by~\eqref{eq:memory-update}, proving~\eqref{eq:main-repair}.
\end{proof}

\begin{corollary}[ODEs for Dual-History Reweighted Repair]
\label{cor:analysis-repair-ode}
Fix the activation fraction $f_0=f$ in the admissible range of
Subsection~\ref{subsec:proof-controls} and take $u\downarrow0$ with
$m\ge m_0\ge1000/u^2$, as required by~\eqref{eq:finite-player-cutoff}. Along repair
steps whose scalar states converge to $(\varsigma,H)$, the
first-order relations on the time scale of a step of length $u$ are
\begin{equation}\label{eq:analysis-repair-ode}
 \dot H=f-H,\qquad \dot\varsigma\ge1+H-\varsigma .
\end{equation}
\end{corollary}

\begin{proof}
The controls~\eqref{eq:finite-retention-controls}
and~\eqref{eq:finite-repair-coefficients} give $t=fu$, $c=1-u+O(u^2)$,
$\beta=fu+O(u^2)$, $1-\rho=(1-f)u+O(u^2)$, and
$\nu=u+O(u^2)$. In particular
$D=\beta+(1-\rho)+\nu H=u(1+H)+O(u^2)$ uniformly on
$0\le H\le L_0$. Also $E/u\le120/m\to0$.
Subtract the current state from~\eqref{eq:main-repair}, divide
by $u$, and pass to the limit.
\end{proof}

\subsection{Derivation of the Potential Function and the Approximation Constant}
\label{subsec:analysis-potential}

Having obtained the drift relations for both branches, we next construct
a potential function to measure their progress. We use the descent
dynamics to derive an explicit expression for this potential, then evaluate
its change under repair to determine the achievable coefficient
through~\eqref{eq:analysis-maxmin}.

\paragraph{Deriving the Potential from Successful LP Steps.}
To derive the potential function, we first consider the scalar limit
with zero reserve $d=0$ and fixed
activation fraction $f_0=f\in(0,1)$. As $u\downarrow0$,
$R_0\to(1-f)/f$ and $Q\to1/f$ by~\eqref{eq:finite-reweighting-controls},
so the descent threshold~\eqref{eq:algorithm-threshold} becomes
$$
S_f(H):=\lim_{u\downarrow0}S_u(H)
=\frac{(1+H)^2}{f+H},
\qquad 0\le H\le f.
$$
By Corollary~\ref{cor:analysis-descent-ode}, the successful-LP
dynamics satisfy $\dot H=-zH$ and
$\dot\varsigma\ge S_f(H)-\varsigma$, where $z\in[0,1]$
is the limiting total LP weight.

Recall that both $\varsigma^+$ and $H^+$ in the successful-LP
guarantee~\eqref{eq:main-descent} contain the same retention
factor $P$. Taking their ratio eliminates this factor,
suggesting a potential of the form
\begin{equation}\label{eq:analysis-potential-form}
\Phi_f(\varsigma,H)=\frac{\varsigma}{H}+F_f(H),
\qquad H>0.
\end{equation}
Our goal is to determine $F_f$ so that~\eqref{eq:analysis-potential-form}
does not decrease along the successful-LP dynamics of
Corollary~\ref{cor:analysis-descent-ode}. By the chain rule,
$$
H\dot\Phi_f
=\dot\varsigma
  +z\bigl(\varsigma-H^2F_f'(H)\bigr)
\ge S_f(H)-\varsigma
  +z\bigl(\varsigma-H^2F_f'(H)\bigr).
$$
The critical case is $z=1$, when the LP uses its entire
direction budget. We choose $F_f$ to make the lower bound
vanish in this case, which leads to
\begin{equation}\label{eq:analysis-synchronization}
H^2F_f'(H)=S_f(H).
\end{equation}
With the choice~\eqref{eq:analysis-synchronization}, the inequality simplifies to
$$
H\dot\Phi_f
\ge(1-z)\bigl(S_f(H)-\varsigma\bigr)\ge0
\qquad\text{whenever }\varsigma\le S_f(H).
$$
Thus the successful-LP dynamics determine the derivative
of $F_f$. To integrate~\eqref{eq:analysis-synchronization}, use the partial-fraction
decomposition
$$
F_f'(H)=\frac{S_f(H)}{H^2}
=\frac{1}{fH^2}
+\frac{2f-1}{f^2H}
+\frac{(1-f)^2}{f^2(H+f)}.
$$
Integrating term by term and choosing a convenient additive
constant gives
$$
\Phi_f(\varsigma,H)
=\frac{\varsigma-1/f}{H}
+\frac{2f-1}{f^2}\log\frac Hf
+\frac{(1-f)^2}{f^2}
  \log\left(1+\frac Hf\right).
$$
In particular, this construction yields the identities
$\partial_\varsigma\Phi_f=1/H$ and
$\partial_H\Phi_f=(S_f(H)-\varsigma)/H^2$.

\paragraph{Determining the Approximation Constant from Repair.}
Next, we examine whether this potential function
increases along the repair dynamics~\eqref{eq:analysis-repair-ode}.
By Corollary~\ref{cor:analysis-repair-ode},
these satisfy $\dot H=f-H$ and
$\dot\varsigma\ge1+H-\varsigma$.
Substituting into the derivatives of $\Phi_f$ gives
$$
H^2\dot\Phi_f
\ge H(1+H-\varsigma)
 +(f-H)(S_f(H)-\varsigma)
=f\bigl(K_f(H)-\varsigma\bigr),
$$
where
\begin{equation}\label{eq:analysis-K}
K_f(H)
:=\frac Hf(1+H)
  +\left(1-\frac Hf\right)S_f(H)
=\frac{(1+H)[f+(2f-1)H]}{f(f+H)}.
\end{equation}
The lower bound on $H^2\dot\Phi_f$ above shows that the repair dynamics increase the potential whenever
$\varsigma<K_f(H)$. Moreover, since
$1+H\le S_f(H)$, we have $K_f(H)\le S_f(H)$
for $0\le H\le f$. Therefore a target below
$\min_{0\le H\le f}K_f(H)$ ensures nonnegative progress
in the successful-LP branch by the descent drift bound
and strictly positive progress in the repair branch
by the repair drift bound just obtained.

For each activation fraction $f$, the worst-case historical weight
determines the certified coefficient $\min_{0\le H\le f}K_f(H)$.
We therefore choose $f$ to maximize this uniform guarantee:
\begin{equation}\label{eq:analysis-maxmin}
\sigma_\star
=\max_{0<f<1}\ \min_{0\le H\le f}K_f(H).
\end{equation}
Solving~\eqref{eq:analysis-maxmin} gives the unique optimal activation
fraction $\fstar\in(7/12,3/5)$ satisfying
$2\fstar^3-4\fstar^2+1=0$. With
$\sstar=1-\fstar$, the optimal value is
$\sigma_\star=9-16\fstar+6\fstar^2$.
The two identities
\begin{equation}\label{eq:analysis-squares}
K_{\fstar}(H)-\sigma_\star
=\frac{(2\fstar-1)(H-\sstar)^2}
        {\fstar(\fstar+H)},\qquad 
\sigma_\star-K_f(\sstar)
=\frac{\sigma_\star(f-\fstar)^2}
        {f(f+\sstar)}
\end{equation}
certify the matching lower and upper bounds when
$f\ge\sstar$. For $f<\sstar$, the boundary value
$K_f(f)=1+f<\sigma_\star$ gives the upper bound.
The complete optimization argument is given in
Theorem~\ref{thm:squares} and
Lemma~\ref{lem:fixed-activation}.
For the finite algorithm, Subsection~\ref{subsec:proof-controls}
selects a rational activation fraction $f_0$ satisfying
\eqref{eq:finite-activation-choice} to preserve a strict reserve
above the target $\sigma$.

Eliminating $\fstar$ yields
$\sigma_\star^3-19\sigma_\star^2+27\sigma_\star+1=0$.
Its unique root in $(1,2)$ is approximately $1.5867$.

\subsection{Finite-Step Corrections and Running Time}
\label{subsec:analysis-finite}

We next turn the first-order calculations into guarantees for the finite
updates of the algorithm. For the fixed target $1<\sigma<\sigma_\star$,
we use the rational controls chosen in Subsection~\ref{subsec:proof-controls}
to account for finite-step errors. The resulting potential
function~\eqref{eq:analysis-finite-potential} yields
the progress and running-time guarantees of Propositions~\ref{prop:potential}
and~\ref{prop:runtime}.
We integrate the finite threshold~\eqref{eq:algorithm-threshold}
through $H^2F_u'(H)=S_u(H)$ and add a slack term $2d/H$,
obtaining
\begin{equation}\label{eq:analysis-finite-potential}
\widehat\Phi_u(\varsigma,H)
=\frac{\varsigma-G_0+2d}{H}
 +(1-a_u)\log\frac H{f_0}
 +a_u\log(1+QH),
\qquad H>0,
\end{equation}
where $a_u$ is defined in~\eqref{eq:finite-potential-controls}.
The additional slack term ensures strict progress despite
finite-step errors, including the repair correction deficit
bounded in~\eqref{eq:finite-error-bound}.

\begin{proposition}[Uniform Progress in Both Branches]
\label{prop:potential}
At every nonterminal state with $H>0$, let
$\Delta\widehat\Phi
=\widehat\Phi_u(\varsigma^+,H^+)
-\widehat\Phi_u(\varsigma,H)$.
A successful LP step satisfies
$H\Delta\widehat\Phi\ge\eta d/2>0$,
whereas a repair step satisfies
$H\Delta\widehat\Phi\ge3d(1-c)>0$.
\end{proposition}

The finite-step proof uses~\eqref{eq:finite-synchronization}, the
finite-parameter analogue of~\eqref{eq:analysis-synchronization}.
The main additional ingredient is the
uniform estimate of Lemma~\ref{lem:finite-uniform}, which controls
the finite repair coefficient even near $H=0$.
The complete proof of Proposition~\ref{prop:potential} is
given in Subsection~\ref{subsec:potential-termination}.

\begin{proposition}[Bounded Iterations and Bit Complexity]
\label{prop:runtime}
For each fixed rational target $\sigma\in(1,\sigma_\star)$,
the number of iterations of Algorithm~\ref{alg:main} is
bounded by a constant depending only on $\sigma$.
Its running time on explicitly listed rational
payoff tensors $\ten{U}_i$ is $B^{O_\sigma(1)}$.
\end{proposition}

Indeed, the descent bound~\eqref{eq:main-descent} gives uniform progress
while $H=0$, and the first repair creates positive historical weight
by~\eqref{eq:algorithm-repair-memory}.
Thereafter, Proposition~\ref{prop:potential} guarantees a
strict increase of the potential at every nonterminal iteration.
Since~\eqref{eq:runtime-potential-range} bounds the potential range, only finitely many
iterations are possible. The precise iteration bound and
rational bit-complexity analysis are deferred to
Subsection~\ref{subsec:runtime}.


\subsection{Combination of the Guarantees}
\label{subsec:main-proof}
With the progress and running-time bounds in place, we can now assemble
the proof of Theorem~\ref{thm:main}. We combine the one-step guarantees
of Propositions~\ref{prop:descent} and~\ref{prop:repair} with the potential
argument to show that the algorithm reaches the target within the stated
running-time bound.

\begin{proof}[Proof of Theorem~\ref{thm:main}]
Fix any rational $\sigma\in(1,\sigma_\star)$ and choose the fixed
rational controls as specified in Subsection~\ref{subsec:proof-controls}.
The scalar identities in~\eqref{eq:analysis-squares}, together with
the uniform finite-step estimate of Lemma~\ref{lem:finite-uniform},
provide a strict reserve above $\sigma$.
In particular the descent margin
$\mu^{(k)}=S_u(H^{(k)})-\sigma^{(k)}-d$
from~\eqref{eq:algorithm-threshold} is positive at every
nonterminal state by~\eqref{eq:joint-margin}.

Lemma~\ref{lem:init} constructs a rational product profile with
$\sigma^{(0)}\ge1$. If $m<m_0$, we return this profile, obtaining
$\Reg(\vct x^{(0)})\le1-1/m$. Assume henceforth that $m\ge m_0$.
The local proofs are used with the joint induction established in
Lemma~\ref{lem:joint-invariant}. Its base case has empty history.
In the successful-LP branch the old certificate masses are multiplied
by the all-base factor $P$ in~\eqref{eq:algorithm-descent-update},
and the historical columns are unchanged.
In the failed-LP branch the old columns and the current dual first
imply floor feasibility~\eqref{eq:joint-floor}; only then is the new
activation law~\eqref{eq:repair-controls} formed.
The corrected retention factor is at least $c$ by
Lemma~\ref{lem:finite-raw-error}, permitting auxiliary certificate
thinning and the addition of the new generation certified
by~\eqref{eq:joint-new-certificate}.
Consequently the exact memory bounds~\eqref{eq:joint-memory-bounds}
and all original historical-column bounds~\eqref{eq:joint-column-bounds}
hold at the next iteration. These arguments verify the
hypotheses of Proposition~\ref{prop:descent} and
Proposition~\ref{prop:repair} at every reachable state.

It remains to establish termination using the bounds in
Subsection~\ref{subsec:runtime}. While $H^{(k)}=0$, every successful
descent increases $\sigma^{(k)}$ by a positive constant depending
only on $\sigma$; the first failed descent instead creates
$H^{(k+1)}=ct>0$ by~\eqref{eq:algorithm-repair-memory}. After positive history is created, it remains
positive. Proposition~\ref{prop:potential} shows that both branches
increase the potential~\eqref{eq:analysis-finite-potential} by a fixed positive amount after dividing
by $H\le L_0$. The potential is bounded above and below on all
subsequent nonterminal states, with range bounded
by~\eqref{eq:runtime-potential-range}, so only a bounded number of further
iterations are possible. The precise bound and the exact rational
bit-complexity argument appear in Proposition~\ref{prop:runtime}.
Thus the algorithm terminates in time $B^{O_\sigma(1)}$.

At termination $\sigma^{(k)}\ge\sigma$, and therefore
$
 \Reg(\vct x^{(k)})
   =1-\frac{\sigma^{(k)}}m
   \le1-\frac\sigma m
$
as required.
\end{proof}


\section{Finite-Step Guarantees of Descent, Repair and Running Time}
\label{sec:detailed-proofs}
In this section, we establish the finite-step guarantees underlying the
preceding analysis. We organize the proof into three modules. We first
verify the parameters
in Subsection~\ref{subsec:proof-controls}, then establish the local descent,
repair, and potential bounds in Subsections~\ref{subsec:successful-descent},
\ref{subsec:failed-repair}, and~\ref{subsec:potential-termination}.
To pass from these local bounds to global validity and termination,
we verify the hypotheses along every run in
Subsection~\ref{subsec:closed-induction}, then establish the
iteration and bit-complexity bounds in Subsection~\ref{subsec:runtime}.

For the remaining calculations, fix an outer iteration $k$ and
write $\varsigma=\sigma^{(k)}$, $\varsigma^+=\sigma^{(k+1)}$,
$H=H^{(k)}$, and $H^+=H^{(k+1)}$ whenever the index is suppressed.

\subsection{Verification of Rational Controls}\label{subsec:proof-controls}
We begin the finite-step analysis by choosing rational controls that
preserve a strict reserve above the target. Guided by the scalar
optimization in Subsection~\ref{subsec:analysis-potential}, we verify
the bounds~\eqref{eq:finite-control-bounds} that will be used throughout
the local estimates.

Fix $L_{\rm col}=9$ and $C_E=120$. Fix $1<\sigma<\sigma_{\star}$.  Since the polynomial defining $\sigma_\star$ in Theorem~\ref{thm:main} is strictly decreasing on $[1,2]$, rational
bisection gives a certified interval for $\sigma_{\star}$; choose a
positive rational $d$ whose interval certificate implies
$\sigma+8d<\sigma_{\star}$.  Choose a rational $f_0\in[7/12,3/5]$ sufficiently close to $f_{\star}$ that
\begin{equation}\label{eq:finite-activation-choice}
\min_{0\le H\le f_0}K_{f_0}(H)>\sigma+7d,
\qquad
\dfrac1{f_0}>\sigma+7d.
\end{equation}
The choice of $f_0$ is effective.  Indeed, Lemma~\ref{lem:fixed-activation}
gives the explicit continuous function $U(f)=\min_{0\le H\le f}K_f(H)$
on $[7/12,3/5]$.  Starting with $N=1$, enumerate the dyadic rationals
$p/2^N$ in this interval and evaluate $U(p/2^N)$ by rational interval
arithmetic (in the middle branch of~\eqref{eq:fixed-activation-value}, isolate the square root by
squaring).  Refine the interval until one candidate satisfies
$$
 U(f_0)>\sigma+7d,\qquad f_0^{-1}>\sigma+7d.
$$
Such a candidate is found because Theorem~\ref{thm:squares} gives
$U(f_\star)=\sigma_\star>\sigma+8d$ and $U$ is continuous.
This gives a terminating rational search. Take the rational
steps $u$ and $\eta$ from~\eqref{eq:algorithm-primary-controls}; since
$L_{\rm col}=9$, they satisfy
$u\le\min\{1/(16L_{\rm col}),d/100\}$ and
$\eta\le\min\{1/4,d/24\}$. Choose the player-count cutoff so that
\begin{equation}\label{eq:finite-player-cutoff}
m_0\ge\left\lceil\max\left\{\dfrac{66}{d},\dfrac{1000}{u^2},
\dfrac{400}{d}\right\}\right\rceil.
\end{equation}

The retention factors and activation intensity used in
\eqref{eq:algorithm-repair-memory} and~\eqref{eq:algorithm-activation} are
\begin{equation}\label{eq:finite-retention-controls}
\begin{aligned}
c&=1-u+\dfrac{u^2}{2}-u^3,\qquad t=f_0u,\\
\rho&=1-(1-f_0)u+\dfrac{(1-f_0)^2u^2}{2}.
\end{aligned}
\end{equation}
The coefficients in the repair gain~\eqref{eq:repair-controls} are
\begin{equation}\label{eq:finite-repair-coefficients}
u_0=u-u^2,\qquad \beta=ct(1+u_0),\qquad \nu=cu_0.
\end{equation}

For the reweighting rule~\eqref{eq:finite-tilt}, we introduce
\begin{equation}\label{eq:finite-reweighting-controls}
R_0=\dfrac{1-\rho}{\beta},\qquad G_0=1+R_0,\qquad Q=\dfrac{\nu}{\beta}.
\end{equation}
For the potential~\eqref{eq:finite-potential-definition} and the
invariant~\eqref{eq:joint-memory-bounds}, define
\begin{equation}\label{eq:finite-potential-controls}
a_u=R_0(Q-1),\qquad L_0=\dfrac{ct}{1-c},\qquad
Q_0=tQ=\dfrac{u_0}{1+u_0}.
\end{equation}
These rational controls satisfy
\begin{equation}\label{eq:finite-control-bounds}
0<R_0<1,\qquad 1<Q<2,\qquad 0<a_u<1,\qquad L_0<f_0,\qquad QL_0<c.
\end{equation}
Define $C_u(H)=R_0(1+H)/(1+QH)$ and write the threshold
in~\eqref{eq:algorithm-threshold} as $S_u(H)=1+H+C_u(H)$.

We now verify~\eqref{eq:finite-control-bounds} from
\eqref{eq:finite-retention-controls}--\eqref{eq:finite-potential-controls}.
Writing $f=f_0$, we have
$$
1-\rho=(1-f)u\left(1-\dfrac{(1-f)u}{2}\right)>0,
\qquad c(1+u_0)\ge1-\dfrac32u^2.
$$
Since $(1-f)/f\le5/7$ and $1-\dfrac32u^2>5/7$,
these inequalities imply $0<R_0<1$.  Moreover
$$
 Q=\dfrac{u_0}{t(1+u_0)}
  =\dfrac{1-u}{f(1+u-u^2)}.
$$
For $u\le1/16$ and $f\in[7/12,3/5]$,
$$
 f(1+u-u^2)\le\dfrac35\dfrac{17}{16}<\dfrac{15}{16}\le1-u,
 \qquad f(1+u-u^2)>\dfrac7{12},
$$
so $1<Q<12/7<2$ and hence $0<a_u=R_0(Q-1)<1$.  Also
$$
(1-c)-cu=\dfrac{u^2}{2}+\dfrac{u^3}{2}+u^4>0,
$$
which gives $L_0=cf_0u/(1-c)<f_0$.  Finally,
$QL_0<c$ is equivalent to $Q_0=tQ<1-c$, and
$$
(1-c)(1+u_0)-u_0
=\dfrac32u^2-\dfrac12u^3+\dfrac32u^4-u^5
=u^2\left(\dfrac32-\dfrac u2+\dfrac{3u^2}{2}-u^3\right)>0
$$
proves this last inequality. These bounds are used in
Lemma~\ref{lem:finite-legal-tilt} for legal reverse allocation and in
Proposition~\ref{prop:potential} for finite-step progress.

On $0\le H\le L_0$,
\begin{equation}\label{eq:finite-threshold-bounds}
 S_u'(H)=1+\dfrac{R_0(1-Q)}{(1+QH)^2}\ge1-R_0(Q-1)>0,\qquad
 S_u(H)\le1+L_0+R_0<\dfrac{13}{5}<3.
\end{equation}
The maintained invariant, proved in Lemma~\ref{lem:joint-invariant},
is $h_i^{(k)}\ge0$, $H^{(k)}=\sum_i h_i^{(k)}\le L_0$, and
$v_i^{(k)}\le R_0H^{(k)}$.

For comparison with the limiting analysis, we also
recall the function from~\eqref{eq:analysis-K}:
$$
K_{f_0}(H)
=
\dfrac{(1+H)[f_0+(2f_0-1)H]}
{f_0(f_0+H)}.
$$

Finally, after each repair iteration, the memory
variables obey the exact update rules in~\eqref{eq:algorithm-repair-memory}.

\subsection{Proofs for Linearized Descent}\label{subsec:successful-descent}

With the controls fixed, we next analyze the linearized LP, whose
outcome determines which update the algorithm takes. We establish
Lemma~\ref{lem:descent} to obtain both the successful-step bound used in
Proposition~\ref{prop:descent} and the dual certificate needed in
Lemma~\ref{lem:finite-legal-tilt} after a failed step.

For $b\in A_j$, the scalar replacement difference is
\(\Delta_{j,b}g_{i,a}(\vct x) =g_{i,a}(\vct x[j\leftarrow\vct e_b])-g_{i,a}(\vct x).\) In the local LP notation below, the profile argument is suppressed when it is clear.
For rational $0<\eta<1$ and $\mu>0$, the epigraph LP
\eqref{eq:linearized-lp} is equivalently
\begin{equation}\label{eq:descent-LP}
 \min_{\vct z\ge\vct0,\;\sum_{j,b}z_{j,b}\le1}
 \max_{i,a}\left[
  g_{i,a}(\vct x)+\eta\sum_{j,b}z_{j,b}\Delta_{j,b}g_{i,a}
 \right].
\end{equation}
This is a rational linear program after introducing an epigraph
variable. It passes when its value is at most
$\Reg(\vct x)-\eta\mu/m$.

\begin{lemma}[Descent and Failed-Dual Columns]\label{lem:descent}
If the linear program passes, put
$$
 Z_j=\sum_bz_{j,b}, Z=\sum_jZ_j\le1, 
 P=\prod_j(1-\eta Z_j),
 \vct x_j^+=(1-\eta Z_j)\vct x_j+\eta\sum_bz_{j,b}\vct e_b.
$$
Then
\begin{equation}\label{eq:descent-bound}
 \varsigma^+\ge P[(1+\eta Z)\varsigma+\eta\mu]
 \ge\varsigma+\eta\mu-\eta^2(\varsigma+\mu).
\end{equation}
Every stored historical mass may be multiplied by $P$.

If the linear program fails, a rational optimal dual gives a player
probability vector $\vct\lambda$ and conditional mixed deviations
$\vct b_i$ such that, for every player $j$ and every
$\vct y_j\in\Delta(A_j)$,
\begin{equation}\label{eq:dual-columns}
 F_j(\vct y_j)
 :=\sum_i\lambda_i
     d_{i,\vct b_i}(\vct x[j\leftarrow\vct y_j])
 <\dfrac{\varsigma+\mu}{m}.
\end{equation}
The same dual satisfies
\begin{equation}\label{eq:dual-base}
 \lambda_j\le F_j(\vct b_j),
 \sum_i\lambda_i d_{i,\vct b_i}(\vct x)
 <\dfrac{\varsigma+\eta\mu}{m}.
\end{equation}
\end{lemma}
\begin{proof}
\emph{Passing Direction.}
For $c=(i,a)$ put $d_c^0=d_{i,a}(\vct x)$ and
$d_c^{j,b}=d_{i,a}(\vct x[j\leftarrow\vct e_b])$.
The product update~\eqref{eq:algorithm-descent-update} samples,
independently for every coordinate, either
the base label (with mass $1-\eta Z_j$) or the label $(j,b)$ (with mass
$\eta z_{j,b}$).  Keeping only the all-base event and the events with
exactly one changed coordinate gives
\[
d_{i,a}(\vct x^+)\ge P\left[d_c^0+
 \sum_{j,b}\dfrac{\eta z_{j,b}}{1-\eta Z_j}d_c^{j,b}\right].
\]
Since $d_c^{j,b}=d_c^0-\Delta_{j,b}g_{i,a}\ge0$ and
$\dfrac{\eta z_{j,b}}{1-\eta Z_j}\ge\eta z_{j,b}$, the difference between
the right-hand side and
\[
P\left[(1+\eta Z)d_c^0-
\eta\sum_{j,b}z_{j,b}\Delta_{j,b}g_{i,a}\right]
\]
is a sum of nonnegative terms
$(\eta z_{j,b}/(1-\eta Z_j)-\eta z_{j,b})d_c^{j,b}$.
The passing inequality says
$d_c^0-\eta\sum_{j,b}z_{j,b}\Delta_{j,b}g_{i,a}
\ge q+\eta\mu/m$, and $d_c^0\ge q$; hence, after multiplying by
$m$ and taking the minimum over $c$,
\[
\varsigma^+\ge P[(1+\eta Z)\varsigma+\eta\mu].
\]
Using $P\ge1-\eta Z$ gives
$$
 P[(1+\eta Z)\varsigma+\eta\mu]
 \ge\varsigma+\eta\mu-\eta^2(Z^2\varsigma+Z\mu)
 \ge\varsigma+\eta\mu-\eta^2(\varsigma+\mu).
$$
The all-base event has probability $P$, which gives the asserted
retention of history in~\eqref{eq:algorithm-descent-memory}.

\emph{Failure Case: Dual Certificate.}
Write~\eqref{eq:descent-LP} in epigraph form. Its exact rational dual is
$$
\begin{aligned}
\max_{w,\alpha}\quad&
 \sum_{i,a}w_{i,a}g_{i,a}(\vct x)-\alpha\\
\text{subject to}\quad&
 \sum_{i,a}w_{i,a}=1,\qquad w_{i,a}\ge0,\quad \alpha\ge0,\\
&\alpha+\eta\sum_{i,a}w_{i,a}\Delta_{j,b}g_{i,a}\ge0
 \quad\text{for every }(j,b).
\end{aligned}
$$
Here $\alpha$ is the multiplier of the constraint
$\sum_{j,b}z_{j,b}\le1$.
Let $w_{i,a}$ be an optimal solution of this dual program and define,
as in~\eqref{eq:algorithm-dual-law},
$$
 \lambda_i=\sum_aw_{i,a},\qquad 
 \vct b_i=\lambda_i^{-1}\sum_aw_{i,a}\vct e_a
       \qquad(\lambda_i>0),
$$
and choose any $\vct b_i$ when $\lambda_i=0$.  Then
$\sum_i\lambda_i=1$.  Each $d_{i,\vct b_i}(\vct x)$ is a convex
combination of the numbers $d_{i,a}(\vct x)\ge q$, so
$\bar q:=\sum_i\lambda_i d_{i,\vct b_i}(\vct x)\ge q$.
The failed test and strong duality give
\[
1-\bar q-\alpha>1-q-\eta\mu/m,
\qquad\text{hence}\qquad
\bar q+\alpha<q+\eta\mu/m.
\]
For the dual constraint indexed by $(j,b)$, use
$\sum_{i,a}w_{i,a}\Delta_{j,b}g_{i,a}
=\bar q-F_j(\vct e_b)$ and obtain
\begin{align*}
 \eta F_j(\vct e_b)
 &\le \alpha+\eta\bar q
 <q-(1-\eta)\bar q+\eta\mu/m
 \le\eta q+\eta\mu/m.
\end{align*}
Linearity in $\vct y_j$ extends this from pure $\vct e_b$ to every
$\vct y_j\in\Delta(A_j)$, proving~\eqref{eq:dual-columns}.
The inequality from the failed LP test and $\alpha\ge0$ give
$\bar q<q+\eta\mu/m$, which is~\eqref{eq:dual-base}.
Finally,
$d_{j,\vct b_j}(\vct x[j\leftarrow\vct b_j])=1$;
all other deficits are nonnegative. Hence
$\lambda_j\le F_j(\vct b_j)$.
\end{proof}

The functions $F_j$ in~\eqref{eq:dual-columns} are weighted replacement
columns: they bound the aggregate deficit after changing one player's
strategy. Lemma~\ref{lem:finite-raw-error} uses these bounds to control
the loss from an additional changed coordinate. The diagonal bound
in~\eqref{eq:dual-base} also controls the mass assigned to any one player.

\begin{proof}[Proof of Proposition~\ref{prop:descent}]
The multilinear expansion over the all-base event and the single-coordinate
events is the passing part of Lemma~\ref{lem:descent}; it gives
\eqref{eq:descent-bound}.  The all-base event retains every historical
record with factor $P^{(k)}$, which proves the update in~\eqref{eq:main-descent}.
\end{proof}

\subsection{Proofs for Reweighted Repair and Historical Certificates}\label{subsec:failed-repair}

Next, we use the dual information from a failed LP to establish the
repair guarantee of Proposition~\ref{prop:repair}. To account for the
contributions of earlier repairs, we construct historical certificates
and use them to control the repair error, leading to the
correction-error bound in Lemma~\ref{lem:finite-raw-error}.

Fix the current outer iteration $k$. For each historical repair event
$s\in\mathcal R^{(k)}$, write $A^{(s,k)}$ while proving the
local estimates below; we also use the current-state convention
$h_i=h_i^{(k)}$ and $H=H^{(k)}$. Thus every sum over $s$ in this
subsection is over $\mathcal R^{(k)}$.

A repair event created at iteration $s$ records its original base profile
$\vct x^{(s)}$, deviations $\vct b_i^{(s)}$, activation law
$\vct\gamma^{(s)}$, intensity $t^{(s)}$, and retained common mass $A^{(s,k)}$.
The mass $A^{(s,k)}$ includes the all-base factors at the repair event's
creation and the retention factors from later updates. Define its
activation odds and the aggregate memory by
\begin{equation}\label{eq:repair-history-representation}
 \alpha_i^{(s)}
 =\dfrac{t^{(s)}\gamma_i^{(s)}}{1-t^{(s)}\gamma_i^{(s)}},\qquad 
 h_i=\sum_{s\in\mathcal R^{(k)}} A^{(s,k)} t^{(s)}\gamma_i^{(s)},\qquad 
 H=\sum_ih_i=\sum_{s\in\mathcal R^{(k)}} A^{(s,k)} t^{(s)}.
\end{equation}
The retained event for player $i$ at generation $s$ is the nested-label
event described formally in Lemma~\ref{lem:certificate-coupling}: it
contains her activation at that generation, the base labels of all other
coordinates at that generation, and every subsequent old/base branch.
The lemma proves that its probability is
$A^{(s,k)}\alpha_i^{(s)}$ and that, conditionally, player $i$ uses
$\vct b_i^{(s)}$ while the opponents have their original base profile.
For a fixed player, retained events from distinct generations are disjoint:
a later activation uses a label that the earlier retained event requires to
be a base label.

The memory $h_i$ in~\eqref{eq:repair-history-representation} uses activation probabilities rather than activation
odds. Since $\alpha_i^{(s)}\ge t^{(s)}\gamma_i^{(s)}$, it is a
convenient lower bound for the retained activation mass.

\begin{lemma}[Cross-Generation Certificate Coupling]
\label{lem:certificate-coupling}
Fix all optimization and tie-breaking choices. Consider a finite prefix of the algorithm ending at the beginning of iteration $k$, with all profiles, mixed deviations, and update coefficients fixed independently of the auxiliary random variables introduced below. Assume that every completed update is legal: it is a well-defined convex mixture, all old/base-branch probabilities are strictly positive, and every repair has $t^{(s)}\gamma_i^{(s)}<1$. Assume also that every completed repair $s<k$ satisfies $\widehat C^{(s)}\ge c>0$, where
$\widehat C^{(s)}=\dfrac{C_{\rm raw}^{(s)}}{1+E^{(s)}}.$
Let $\mathcal R^{(k)}$ be the set of completed repair indices, and put
$$\alpha_i^{(s)}:=\frac{t^{(s)}\gamma_i^{(s)}}{1-t^{(s)}\gamma_i^{(s)}}.$$
Define the common scalar coefficients by $A^{(s,s+1)}=c$. For each $s\in\mathcal R^{(\ell)}$, a descent at iteration $\ell$ gives
$$A^{(s,\ell+1)}=P^{(\ell)}A^{(s,\ell)},$$
whereas a repair at iteration $\ell$ gives
$$A^{(s,\ell+1)}=cA^{(s,\ell)}.$$
For a mixed profile $\vct z$, write $\mathbb E_{\vct z}[\varphi]$ for the expectation of $\varphi$ under the product law $\bigotimes_{\ell=1}^m\vct z_\ell$.

There exist an augmented probability space $(\Omega_k,\mathbb P_k)$, an action map $X^{(k)}$ with law $\bigotimes_i\vct x_i^{(k)}$, and certificate events $E_{s,i}^{(k)}$, for $s\in\mathcal R^{(k)}$ and $i\in[m]$, with the following properties and one-step extensions.

\begin{enumerate}[label=\textup{(C\arabic*)},leftmargin=2.6em]
\item\label{item:certificate-mass}
The event masses are
$$\mathbb P_k(E_{s,i}^{(k)})=A^{(s,k)}\alpha_i^{(s)}.$$
\item\label{item:certificate-override}
For every bounded function $\varphi$ on $\prod_{\ell=1}^m A_\ell$,
$$\mathbb E_k\!\left[\mathbf 1_{E_{s,i}^{(k)}}\varphi(X^{(k)})\right]=A^{(s,k)}\alpha_i^{(s)}\mathbb E_{\vct x^{(s)}[i\leftarrow\vct b_i^{(s)}]}[\varphi].$$
Thus, whenever the event has positive probability, its conditional action law is the generation-$s$ product profile with player $i$ using $\vct b_i^{(s)}$.
For any fixed $j$ and $\vct y_j\in\Delta(A_j)$, enlarge the space by an independent action $Y_j\sim\vct y_j$ and define, on the entire space,
$$X_j^{(k;j\leftarrow Y_j)}:=Y_j,\qquad X_\ell^{(k;j\leftarrow Y_j)}:=X_\ell^{(k)}\quad(\ell\ne j).$$
Keep every branch label and every certificate event unchanged. This map has the product law $\vct x^{(k)}[j\leftarrow\vct y_j]$. Moreover, simultaneously for every $s$ and every $i\ne j$,
$$\mathbb E\!\left[\mathbf 1_{E_{s,i}^{(k)}}\varphi(X^{(k;j\leftarrow Y_j)})\right]=A^{(s,k)}\alpha_i^{(s)}\mathbb E_{\vct x^{(s)}[i\leftarrow\vct b_i^{(s)},\,j\leftarrow\vct y_j]}[\varphi].$$
Both identities include zero-probability certificate events.
\item\label{item:certificate-descent}
A legal descent transition~\eqref{eq:algorithm-descent-update} extends the construction by intersecting every old certificate event with an all-base cylinder independent of $\Omega_k$ and of probability $P^{(k)}$. The event masses are multiplied by $P^{(k)}$, and their generation-specific product laws in \textup{(C2)} are preserved.
\item\label{item:certificate-repair}
For any further legal repair, the raw and corrected action maps and pairwise disjoint pre-thinning events can be constructed without assuming $\widehat C^{(k)}\ge c$. The raw action map has product law $\widetilde{\vct x}^{(k)}$, and the corrected map has product law $\vct x^{(k+1)}$. The pre-thinning old and new event masses are, respectively,
$$\mathbb P(\widetilde E_{s,i}^{(k+1)})=\widehat C^{(k)}A^{(s,k)}\alpha_i^{(s)},\qquad 
\mathbb P(\widetilde E_{k,i}^{(k+1)})=\widehat C^{(k)}\alpha_i^{(k)}.$$
Their action laws satisfy the unnormalized identities in \textup{(C2)} with these masses: old events retain their original generation laws, and a new event uses $\vct x^{(k)}[i\leftarrow\vct b_i^{(k)}]$. Here $\widehat C^{(k)}$ is a common scalar factor in the event masses.
\item\label{item:certificate-thinning}
If the further repair satisfies $\widehat C^{(k)}\ge c$, attach to each pre-thinning event a fresh Bernoulli coin with success probability $c/\widehat C^{(k)}$. The coins are mutually independent and independent of the entire pre-thinning probability space. Intersect each event with its coin-one event. The resulting certificate masses are
$$\mathbb P_{k+1}(E_{s,i}^{(k+1)})=cA^{(s,k)}\alpha_i^{(s)},\qquad 
\mathbb P_{k+1}(E_{k,i}^{(k+1)})=c\alpha_i^{(k)}.$$
The action map is unchanged, and the unnormalized generation-law identities are preserved. Thus the common coefficients update as
$$A^{(s,k+1)}=cA^{(s,k)},\qquad A^{(k,k+1)}=c.$$
\item\label{item:certificate-disjointness}
The certificate events are pairwise disjoint over all pairs $(s,i)$. Disjointness is preserved by descent, holds for all pre-thinning repair events, and is preserved by thinning.
\end{enumerate}

The assumption $\widehat C^{(k)}\ge c$ is needed only for the thinning in \textup{(C5)}, not for the one-step construction in \textup{(C4)}.
\end{lemma}

\begin{proof}
We prove the assertion for a finite prefix by induction on its length, and simultaneously construct the stated one-step extensions. All algorithmic quantities are fixed throughout this probability-space construction. In particular, the pure best responses in both sweeps are computed from their respective intermediate mixed profiles, before any auxiliary sampling.

At iteration $0$, sample the coordinates of $X^{(0)}$ independently with laws $\vct x_i^{(0)}$. Since $\mathcal R^{(0)}=\varnothing$, there are no certificate events. Suppose that the construction is available at iteration $k$. We identify every old event with its pullback to each subsequent product extension.

\emph{Descent.}
For each player $j$, sample a label with mutually exclusive outcomes $O_j$ and $N_{j,b}$, $b\in A_j$, having probabilities
$$\mathbb P(O_j)=1-\eta Z_j^{(k)},\qquad \mathbb P(N_{j,b})=\eta z_{j,b}^{(k)}.$$
These labels are independent across players and independent of the entire old space. Set $X_j^{(k+1)}=X_j^{(k)}$ on $O_j$ and $X_j^{(k+1)}=b$ on $N_{j,b}$. This gives exactly the product law prescribed by \eqref{eq:algorithm-descent-update}. Let
$$B_D^{(k)}:=\bigcap_j O_j,\qquad E_{s,i}^{(k+1)}:=E_{s,i}^{(k)}\cap B_D^{(k)}.$$
The cylinder $B_D^{(k)}$ has probability $P^{(k)}$, is independent of the old space, and forces $X^{(k+1)}=X^{(k)}$. Hence
$$\mathbb E\!\left[\mathbf 1_{E_{s,i}^{(k+1)}}\varphi(X^{(k+1)})\right]=P^{(k)}\mathbb E_k\!\left[\mathbf 1_{E_{s,i}^{(k)}}\varphi(X^{(k)})\right].$$
The induction hypothesis proves the required identity with $A^{(s,k+1)}=P^{(k)}A^{(s,k)}$. Intersecting disjoint events with the same cylinder preserves disjointness.

\emph{Repair before thinning.}
Suppress the superscript $(k)$ on the current repair parameters. Let $a_j^{\rm rev}$ and $a_j^{\rm cor}$ be the fixed pure actions selected by the reverse and forward sweeps, respectively. The corresponding mixing probabilities are
$$p_j=\frac{\delta_j}{P_{j-1}},\qquad \theta_j=\frac{e_j}{1+\sum_{\ell\le j}e_\ell},$$
where the correction deficits $e_\ell$ are frozen before the forward sweep. For every player $j$, sample
$$R_j\sim{\rm Bernoulli}(p_j),\qquad C_j\sim{\rm Bernoulli}(\theta_j),\qquad V_j\sim\vct b_j^{(k)},$$
and a label $L_j\in\{B,A\}$ with
$$\mathbb P(L_j=A)=t\gamma_j,\qquad \mathbb P(L_j=B)=1-t\gamma_j.$$
All these random variables are mutually independent and independent of the entire old space, including all earlier certificate coins. In particular, $L_j$ and $V_j$ are sampled even when $R_j=1$; their values are then ignored by the raw action map.
Define the raw action map as follows. On $\{R_j=1\}$, set $\widetilde X_j=a_j^{\rm rev}$; on $\{R_j=0,L_j=A\}$, set $\widetilde X_j=V_j$; and on $\{R_j=0,L_j=B\}$, set $\widetilde X_j=X_j^{(k)}$. Define the corrected map by setting $X_j^{(k+1)}=a_j^{\rm cor}$ on $\{C_j=1\}$ and $X_j^{(k+1)}=\widetilde X_j$ on $\{C_j=0\}$.
Their coordinate marginals are exactly those prescribed by \eqref{eq:algorithm-activation}, \eqref{eq:algorithm-reverse-sweep}, and \eqref{eq:algorithm-forward-correction}. Since each coordinate uses only its old action and its own fresh independent variables, both maps have the required product laws.

Put
$$B_R^{(k)}:=\bigcap_j\{R_j=0,L_j=B\},\qquad B_C^{(k)}:=\bigcap_j\{C_j=0\}.$$
The reverse-sweep identity gives
$$\mathbb P(B_R^{(k)})=\prod_j(1-p_j)(1-t\gamma_j)=\rho\prod_j(1-t\gamma_j)=C_{\rm raw}^{(k)}.$$
The correction identity gives
$$\mathbb P(B_C^{(k)})=\prod_j(1-\theta_j)=\frac{1}{1+E^{(k)}}.$$
Consequently,
$$\mathbb P(B_R^{(k)}\cap B_C^{(k)})=\widehat C^{(k)}.$$
For old generations define
$$\widetilde E_{s,i}^{(k+1)}:=E_{s,i}^{(k)}\cap B_R^{(k)}\cap B_C^{(k)},\qquad s\in\mathcal R^{(k)},$$
and define the new-generation events by
$$\widetilde E_{k,i}^{(k+1)}:=B_C^{(k)}\cap\{R_i=0,L_i=A\}\cap\bigcap_{j\ne i}\{R_j=0,L_j=B\}.$$
The all-base cylinders are independent of the old space and force the corrected map to equal the old map. Consequently,
$$\mathbb E\!\left[\mathbf 1_{\widetilde E_{s,i}^{(k+1)}}\varphi(X^{(k+1)})\right]=\widehat C^{(k)}A^{(s,k)}\alpha_i^{(s)}\mathbb E_{\vct x^{(s)}[i\leftarrow\vct b_i^{(s)}]}[\varphi].$$
For a new-generation event, independence gives
$$\mathbb P(\widetilde E_{k,i}^{(k+1)})=\prod_j(1-\theta_j)\,(1-p_i)t\gamma_i\prod_{j\ne i}(1-p_j)(1-t\gamma_j).$$
Using $t\gamma_i<1$, this becomes
$$\mathbb P(\widetilde E_{k,i}^{(k+1)})=\widehat C^{(k)}\frac{t\gamma_i}{1-t\gamma_i}=\widehat C^{(k)}\alpha_i^{(k)}.$$
Its defining event depends only on the fresh branch labels, independently of $X^{(k)}$ and all the $V_j$. On this event the corrected map is $(V_i,X_{-i}^{(k)})$, in coordinate order. Therefore
$$\mathbb E\!\left[\mathbf 1_{\widetilde E_{k,i}^{(k+1)}}\varphi(X^{(k+1)})\right]=\widehat C^{(k)}\alpha_i^{(k)}\mathbb E_{\vct x^{(k)}[i\leftarrow\vct b_i^{(k)}]}[\varphi].$$
These identities also hold when $\gamma_i=0$, without conditioning on a null event.

The old pre-thinning events remain disjoint by induction. Each old event requires all current activation labels to be $B$, whereas a new event requires $L_i=A$ at its target coordinate. Thus old and new events are disjoint. Two new events with different targets are disjoint because one requires $L_i=A$ and the other requires $L_i=B$. This establishes \textup{(C4)} for every legal repair, without any assumption that $\widehat C^{(k)}\ge c$.

\emph{Thinning.}
If $\widehat C^{(k)}\ge c$, enlarge the pre-thinning space by mutually independent coins
$$T_{s,i}\sim{\rm Bernoulli}\!\left(\frac{c}{\widehat C^{(k)}}\right),\qquad s\in\mathcal R^{(k)}\cup\{k\},\quad i\in[m],$$
independent of its entire sigma-field, and set
$E_{s,i}^{(k+1)}:=\widetilde E_{s,i}^{(k+1)}\cap\{T_{s,i}=1\}.$
The action map does not use these coins. For every such index pair,
$$\mathbb E\!\left[\mathbf 1_{E_{s,i}^{(k+1)}}\varphi(X^{(k+1)})\right]=\frac{c}{\widehat C^{(k)}}\mathbb E\!\left[\mathbf 1_{\widetilde E_{s,i}^{(k+1)}}\varphi(X^{(k+1)})\right].$$
The preceding identities prove \textup{(C1)} and the first identity in \textup{(C2)} with the stated updates of $A$. Taking subsets preserves disjointness, and the product action law is unchanged. This completes the induction for all completed updates.
\emph{Global overrides.}
Fix $j$ and $\vct y_j$. On a further product extension, sample $Y_j\sim\vct y_j$ independently of the entire space and use the global replacement map defined in \textup{(C2)}. Its unconditional law is the required product law because $X_{-j}^{(k)}$ has the corresponding product marginal and is independent of $Y_j$. All old labels and certificate events are retained on this extension.

For a bounded $\varphi$, define
$$g(z):=\sum_{a\in A_j}(\vct y_j)_a\,\varphi(z[j\leftarrow a]).$$
Independence of $Y_j$ gives
$\mathbb E\!\left[\mathbf 1_{E_{s,i}^{(k)}}\varphi(X^{(k;j\leftarrow Y_j)})\right]=\mathbb E_k\!\left[\mathbf 1_{E_{s,i}^{(k)}}g(X^{(k)})\right].$
Applying the first identity in \textup{(C2)} to $g$ yields
$$\mathbb E_k\!\left[\mathbf 1_{E_{s,i}^{(k)}}g(X^{(k)})\right]=A^{(s,k)}\alpha_i^{(s)}\mathbb E_{\vct x^{(s)}[i\leftarrow\vct b_i^{(s)}]}[g].$$
For $i\ne j$, the definition of $g$ implies
$$\mathbb E_{\vct x^{(s)}[i\leftarrow\vct b_i^{(s)}]}[g]=\mathbb E_{\vct x^{(s)}[i\leftarrow\vct b_i^{(s)},\,j\leftarrow\vct y_j]}[\varphi].$$
Combining these equalities proves the override identity in \textup{(C2)}. The same global map works simultaneously for all these certificate events. Taking $\vct y_j=\boldsymbol e_b$ includes pure overrides. The argument also applies to the pre-thinning identities above. No step requires conditioning on a zero-probability event.
\end{proof}

The strategy trees and certificate coins serve as witnesses for the analysis.
The algorithm manipulates rational product marginals and fixed best-response
labels deterministically.

By property~\ref{item:certificate-override} of Lemma~\ref{lem:certificate-coupling},
these conditional laws remain valid when coordinate $j\ne i$ is
overridden by any mixed strategy $\vct y_j$, because its auxiliary
labels can still be sampled. Define the original columns by
\begin{equation}\label{eq:repair-original-columns}
d_{ij}^{(s)}(\vct y_j) =d_{i,\vct b_i^{(s)}} (\vct x^{(s)}[j\leftarrow\vct y_j]).
\end{equation}
A historical column certificate is the inequality
\begin{equation}\label{eq:historical-column-certificate}
 \sum_{i\ne j}\gamma_i^{(s)}d_{ij}^{(s)}(\vct y_j)
 \le L_{\rm col}/m
 \qquad\text{for every }j,\vct y_j.
\end{equation}
The column remains evaluated at its original base profile and deviations
after later updates; its certificate bounds the aggregate loss from
overriding one coordinate of a retained event.

For a fresh failed dual~\eqref{eq:algorithm-dual-law}, put
$$
 \pi_i=m\lambda_i,\qquad
 r=\max_i\pi_i-1,\qquad 
 v_i=mh_i-H,\qquad 
 H\psi=\sum_i\lambda_i v_i.
$$
Only the product $H\psi$ is needed at $H=0$. In particular,
$m\sum_i\lambda_ih_i=H(1+\psi)$.

\begin{lemma}[Finite Historical Overlap]\label{lem:history-overlap}
Fix a failed descent LP at the current state, and let
$(\vct\lambda,(\vct b_i)_i)$ be an optimal rational dual
certificate satisfying~\eqref{eq:dual-columns}.
Assume $h_i\ge0$, $H=\sum_i h_i$, $Y\ge0$, and
$mh_i\le H+Y$ for all $i$. Suppose that every stored repair
generation satisfies~\eqref{eq:historical-column-certificate}
and has the retained event law of Lemma~\ref{lem:certificate-coupling}.
Then
\begin{equation}\label{eq:historical-overlap}
 (1+r)(1-\vartheta)+H(1+\psi)<\varsigma+\mu,\qquad 
 0\le\vartheta\le\dfrac{H+Y+L_{\rm col}H}{m}.
\end{equation}
\end{lemma}
\begin{proof}
By Lemma~\ref{lem:certificate-coupling}, the retained events are
pairwise disjoint cylinder events with the stated conditional laws. Hence,
for $i\ne j$, their nonnegative payoff contributions give
\(u_i(\vct x[j\leftarrow\vct y_j]) \ge\sum_{s\in\mathcal R^{(k)}} A^{(s,k)} t^{(s)}\gamma_i^{(s)} [1-d_{ij}^{(s)}(\vct y_j)].\)
Indeed, the conditional payoff is at least the bracketed expression.
For clarity, put
$z_s=1-d_{ij}^{(s)}(\vct y_j)$ and
$w_s=A^{(s,k)}\alpha_i^{(s)}$.  The disjoint-event decomposition gives
$u_i(\vct x[j\leftarrow\vct y_j])\ge\sum_s w_s(z_s)_+$.
For $z_s\ge0$ we have
$w_s\ge A^{(s,k)}t^{(s)}\gamma_i^{(s)}$, while for $z_s<0$ the term
$A^{(s,k)}t^{(s)}\gamma_i^{(s)}z_s$ is nonpositive.  Therefore
\[
 u_i(\vct x[j\leftarrow\vct y_j])
 \ge\sum_s A^{(s,k)}t^{(s)}\gamma_i^{(s)}z_s
 =\sum_s A^{(s,k)}t^{(s)}\gamma_i^{(s)}
       [1-d_{ij}^{(s)}(\vct y_j)].
\]
Every current fixed-deviation deficit dominates the actual payoff because
$d_{i,\vct b_i}(z)=1-u_i(z[i\leftarrow\vct b_i])+u_i(z)\ge u_i(z)$.
At $\vct y_j=\vct b_j$ its diagonal value is one, so
$$
 F_j(\vct b_j)\ge
 \lambda_j+\sum_i\lambda_ih_i-\lambda_jh_j
-\sum_{s\in\mathcal R^{(k)}} A^{(s,k)} t^{(s)}\sum_{i\ne j}
    \lambda_i\gamma_i^{(s)}d_{ij}^{(s)}(\vct b_j).
$$
Choose $j$ with $\lambda_j=(1+r)/m$.  Since
$mh_j\le H+Y$,
\(
 \lambda_jh_j\le\dfrac{(1+r)(H+Y)}{m^2}.
\)
Moreover, using $\lambda_i\le(1+r)/m$ and the column certificate
\eqref{eq:historical-column-certificate},
\[
\sum_s A^{(s,k)}t^{(s)}
  \sum_{i\ne j}\lambda_i\gamma_i^{(s)}
       d_{ij}^{(s)}(\vct b_j)
\le
 \dfrac{1+r}{m}\sum_s A^{(s,k)}t^{(s)}
 \dfrac{L_{\rm col}}m
 =\dfrac{(1+r)L_{\rm col}H}{m^2}.
\]
Finally,
$m\sum_i\lambda_i h_i=H(1+\psi)$ by the definition of $H\psi$.
Thus the preceding lower bound on
$F_j(\vct b_j)$ becomes
\[
 mF_j(\vct b_j)\ge
 (1+r)+H(1+\psi)
 -\dfrac{(1+r)(H+Y+L_{\rm col}H)}m
 =(1+r)(1-\vartheta)+H(1+\psi).
\]
Combining this with $mF_j(\vct b_j)<\varsigma+\mu$ from
\eqref{eq:dual-columns} proves~\eqref{eq:historical-overlap}, with
$\vartheta=(H+Y+L_{\rm col}H)/m$.
\end{proof}

\paragraph{Activation and the Reverse Sweep.}
We analyze the update in~\eqref{eq:algorithm-activation}
and~\eqref{eq:algorithm-reverse-sweep}. Fix the actual activation law $\vct\gamma$, with
$0\le t\gamma_i<1$, and form
\(\vct y_i=(1-t\gamma_i)\vct x_i+t\gamma_i\vct b_i.\)
Let $\delta_i\ge0$ and $\sum_i\delta_i=1-\rho<1$, and set
$P_i=1-\sum_{j\le i}\delta_j$, with $P_0=1$.
In the order $i=m,m-1,\ldots,1$, compute a pure best response
$\vct e_i^{\rm rev}$ to $\vct y_j$ for $j<i$ and the already
constructed raw strategies for $j>i$, and define
\(\widetilde{\vct x}_i =\left(1-\dfrac{\delta_i}{P_{i-1}}\right)\vct y_i +\dfrac{\delta_i}{P_{i-1}}\vct e_i^{\rm rev}.\)
All mixing probabilities are fixed before the sweep; only the pure
best responses are chosen sequentially.

\paragraph{Event Probabilities and Odds.}
The all-base probability from~\eqref{eq:algorithm-correction} is
\(C_{\rm raw}=\rho\prod_i(1-t\gamma_i).\)
Relative to this event, the activation and reverse-change odds are
\begin{equation}\label{eq:repair-event-odds}
\alpha_i=\dfrac{t\gamma_i}{1-t\gamma_i},\qquad
\omega_i=\dfrac{\delta_i}{P_i},\qquad S_{\rm ev}=\sum_i(\alpha_i+\omega_i).
\end{equation}
Their sum $S_{\rm ev}$ measures the available single-change memory.
Since $\sum_i\alpha_i\ge t$ and
$\prod_i(1+\omega_i)=1/\rho$,
\begin{equation}\label{eq:event-memory}
 S_{\rm ev}\ge t+\log(1/\rho).
\end{equation}

To obtain the event decomposition in~\eqref{eq:rawrepair}, sample the
auxiliary labels as follows. Independently for every player, first sample a reverse
label $R_i$ with probability $p_i=\delta_i/P_{i-1}$.  Independently sample a
latent current label $L_i\in\{A_i,B_i\}$ with probabilities
$t\gamma_i$ and $1-t\gamma_i$.  The realized current action uses the
reverse best response when $R_i=1$ and otherwise uses $A_i$ or $B_i$
according to $L_i$.  Thus the all-base event is $R_i=0,B_i$ for every
$i$ and has probability
\[
 \prod_i(1-p_i)(1-t\gamma_i)=\rho\prod_i(1-t\gamma_i)=C_{\rm raw}.
\]
Relative to it, the single-player events $R_i=0,A_i$ and $R_i=1,B_i$
have odds
\[
 \alpha_i=\dfrac{t\gamma_i}{1-t\gamma_i},
 \qquad \omega_i=\dfrac{p_i}{1-p_i}=\dfrac{\delta_i}{P_i},
\]
respectively.  The reverse best-response memory itself uses the whole
event $R_i=1$ (with either latent label), which, together with the
required lower-index base reverse labels, has probability exactly
$\dfrac{\delta_i}{P_{i-1}}P_{i-1}=\delta_i$; the latent label is only used
to select disjoint singleton reverse-change subevents in the other memory
terms.  Labels at different players are independent.  The reverse best
responses $\vct e_i^{\rm rev}$ are computed during the sweep and are fixed
pure actions on this probability space.

A retained activation from generation $s$, as constructed in
Lemma~\ref{lem:certificate-coupling}, requires its activation label at
$s$, base labels for all other players at that generation, and base labels
at every later generation.  Consequently retained events from different
generations are disjoint: at the later generation, the earlier retained
event requires all base labels, whereas the newly created event requires
one activation label.  The four event families in Lemma~\ref{lem:rawrepair} are
therefore disjoint after the indicated exactly-one-opponent-label
restriction; activation and reverse labels at the same player are never
counted as independent events.

\begin{lemma}[Raw Repair and Corrected Retention]\label{lem:rawrepair}
Let
$$
 d_i^0=d_{i,\vct b_i}(\vct x),\qquad 
 d_{ij}(\vct y_j)=d_{i,\vct b_i}(\vct x[j\leftarrow\vct y_j]).
$$
The raw profile satisfies
\begin{equation}\label{eq:rawrepair}
 q_i(\widetilde{\vct x})\ge{}
 C_{\rm raw}q+\delta_i
 +C_{\rm raw}\alpha_i(1+S_{\rm ev})+C_{\rm raw}S_{\rm ev}\sum_{s\in\mathcal R^{(k)}} A^{(s,k)}\alpha_i^{(s)}
 -e_i^{\rm new}-\sum_{s\in\mathcal R^{(k)}} e_i^{(s)},
\end{equation}
where the current-generation error is
\begin{equation}\label{eq:raw-current-error}
 e_i^{\rm new}=C_{\rm raw}\alpha_i
 \biggl[d_i^0+\alpha_i+\omega_i+\sum_{j\ne i}\alpha_jd_{ij}(\vct b_j)+\sum_{j\ne i}\omega_jd_{ij}(\vct e_j^{\rm rev})\biggr],
\end{equation}
and the error from a stored repair event $s$ is
\begin{equation}\label{eq:raw-history-error}
 e_i^{(s)}=C_{\rm raw}A^{(s,k)}\alpha_i^{(s)}
 \biggl[\alpha_i+\omega_i
 +\sum_{j\ne i}\alpha_jd_{ij}^{(s)}(\vct b_j)
 +\sum_{j\ne i}\omega_jd_{ij}^{(s)}(\vct e_j^{\rm rev})\biggr].
\end{equation}
Suppose the memory weights satisfy
\(C_{\rm raw}t(1+S_{\rm ev})\ge\beta, C_{\rm raw}S_{\rm ev}\ge\nu,\)
and the reverse budgets are nonnegative with
\(D=1-\rho+\beta+\nu H, \delta_i=D/m-\beta\gamma_i-\nu h_i.\)
Define, before any further strategy changes,
\(e_i^{\rm cor} =\left(C_{\rm raw}q+D/m-q_i(\widetilde{\vct x})\right)_+, E=\sum_ie_i^{\rm cor}.\)
Then
\(E\le\sum_i\left(e_i^{\rm new}+\sum_{s\in\mathcal R^{(k)}} e_i^{(s)}\right),\)
and the rational forward correction~\eqref{eq:algorithm-forward-correction} gives
\begin{equation}\label{eq:corrected-bound}
 \varsigma^+\ge\dfrac{C_{\rm raw}\varsigma+D}{1+E},
 \qquad \widehat C=\dfrac{C_{\rm raw}}{1+E}.
\end{equation}
If $\widehat C\ge c$, certificate thinning in property~\ref{item:certificate-thinning}
of Lemma~\ref{lem:certificate-coupling} gives
\begin{equation}\label{eq:memory-update}
 H^+=c(H+t),
 \qquad \vct h^+=c(\vct h+t\vct\gamma),
\end{equation}
and $\varsigma^+\ge c\varsigma+D-DE$.
\end{lemma}
\begin{proof}
\emph{Stage 1: Information From Disjoint Events.}
All events below are cylinder events in the augmented space of
Lemma~\ref{lem:certificate-coupling}. For each fixed pure deviation $a$,
$d_{i,a}$ is affine in the event mixture, so we first prove the lower
bound in~\eqref{eq:rawrepair} for every $a$ and only then take $\min_a$. If a current
activation uses $\vct b_i$ at a profile $z$, then
\[
 d_{i,a}(z[i\leftarrow\vct b_i])
 =1-u_i(z[i\leftarrow\boldsymbol e_a])+u_i(z[i\leftarrow\vct b_i])
 \ge u_i(z[i\leftarrow\vct b_i])
 \ge 1-d_{i,\vct b_i}(z),
\]
using $u_i\le1$ and $u_i(z)\ge0$. A reverse best response gives
$d_{i,a}\ge1$ for every $a$. The same pointwise inequalities apply to
historical activation events with the original column~\eqref{eq:repair-original-columns}.
Fix a player $i$ and retain the following four event families.
\begin{enumerate}[label=\textup{(\roman*)},leftmargin=2em,
  itemsep=0.45em,topsep=0.4em]
\item\label{item:raw-all-base} \emph{All Base Labels.}
The event that all current labels are base labels has probability
$C_{\rm raw}$ and contributes at least $C_{\rm raw}q$.

\item\label{item:raw-reverse-response} \emph{The Reverse Best Response.}
Player $i$ uses her reverse label, and every $j<i$ uses its base
reverse label. Labels at $j>i$ are unrestricted. The probability is
\(\dfrac{\delta_i}{P_{i-1}}\prod_{j<i}\dfrac{P_j}{P_{j-1}} =\delta_i,\)
and the conditional regret is zero; hence every fixed-deviation deficit is at least $1$.

\item\label{item:raw-current-activation} \emph{The Current Activation.}
Player $i$ uses her current activation and her base reverse label.
Among opponents, either no label changes or exactly one label changes.
In the latter case, the changed label is either an opponent's
activation $(R_j=0,L_j=A)$ or its \emph{singleton reverse-change} label
$(R_j=1,L_j=B)$; all other current labels are base labels.
The full reverse event with $L_j$ unrestricted is used only for the
best-response memory in case~\ref{item:raw-reverse-response}. These events contribute at least
\[
 C_{\rm raw}\alpha_i\biggl[1-d_i^0
 +\sum_{j\ne i}\alpha_j(1-d_{ij}(\vct b_j))+\sum_{j\ne i}\omega_j
       (1-d_{ij}(\vct e_j^{\rm rev}))\biggr].
\]

\item\label{item:raw-retained-activation} \emph{A Retained Activation.}
Player $i$ has her retained activation event from repair event $s$,
both of her current labels are base labels, and exactly one current
opponent label changes, using the same singleton reverse-change event
$(R_j=1,L_j=B)$ whenever the changed label is a reverse one.
This generation contributes at least
\[
 C_{\rm raw}A^{(s,k)}\alpha_i^{(s)}\biggl[\sum_{j\ne i}\alpha_j
       (1-d_{ij}^{(s)}(\vct b_j))+\sum_{j\ne i}\omega_j
       (1-d_{ij}^{(s)}(\vct e_j^{\rm rev}))\biggr].
\]
Overridden opponent labels remain sampled, so the original conditional
laws defining $d_{ij}^{(s)}$ are preserved.
\end{enumerate}
The families in cases~\ref{item:raw-all-base}--\ref{item:raw-retained-activation}
are disjoint. The second uses player $i$'s reverse
label, the third uses her activation and base reverse label, and the
fourth keeps both of her current labels at base. An opponent change
separates the fourth family from the all-base event, and its retained
events from distinct repair events are also disjoint.
More explicitly, write
$B_j=\{R_j=0,L_j=B\}$,
$A_j=\{R_j=0,L_j=A\}$,
$V_j^{\rm one}=\{R_j=1,L_j=B\}$, and
$V_j^{\rm all}=\{R_j=1\}$.  The event $V_j^{\rm one}$, and \emph{not}
$V_j^{\rm all}$, has the single-change odds in~\eqref{eq:repair-event-odds}:
\[
 \dfrac{\mathbb P(V_j^{\rm one})}{\mathbb P(B_j)}
 =\dfrac{p_j(1-t\gamma_j)}{(1-p_j)(1-t\gamma_j)}
 =\dfrac{p_j}{1-p_j}=\omega_j.
\]
The full reverse best-response event for player $i$ is
$V_i^{\rm all}\cap\bigcap_{j<i}\{R_j=0\}$ and has mass $\delta_i$;
it deliberately leaves $L_i$ unrestricted.
The current-activation singletons have signatures
$A_i\cap\bigcap_{j\ne i}B_j$,
$A_i\cap A_j\cap\bigcap_{\ell\notin\{i,j\}}B_\ell$, or
$A_i\cap V_j^{\rm one}\cap\bigcap_{\ell\notin\{i,j\}}B_\ell$
for $j\ne i$.  A retained activation from generation $s$ has its
generation-$s$ activation label, current $B_i$, and exactly one current
opponent label $A_j$ or $V_j^{\rm one}$, with all other current labels $B$.
These signatures are pairwise disjoint: within the current generation
they prescribe different labels at the target coordinate, while for
distinct historical generations the earlier retained event prescribes
base labels at the later generation and the later event prescribes an
activation label there. The lower bound~\eqref{eq:rawrepair} keeps only the
retained subevents with exactly one current opponent label change.
The no-change retained subevents are omitted because every fixed
deficit is nonnegative, so omission can only weaken the lower bound.
The same disjointness remains valid under a pure or mixed opponent
override, since the override changes only the action leaf and not the
label signature.

Expand $S_{\rm ev}=\sum_j(\alpha_j+\omega_j)$ in the contributions
from cases~\ref{item:raw-current-activation} and~\ref{item:raw-retained-activation}.
Only opponent changes receive single-change memory, so the terms
$\alpha_i+\omega_i$ enter the errors. This gives
\eqref{eq:rawrepair}--\eqref{eq:raw-history-error}.

\emph{Stage 2: Balancing the memory.} For every $i$, the positive
terms in \eqref{eq:rawrepair} satisfy
\begin{align*}
 C_{\rm raw}\alpha_i(1+S_{\rm ev})
 &\ge C_{\rm raw}t\gamma_i(1+S_{\rm ev})\ge\beta\gamma_i,\\
 C_{\rm raw}S_{\rm ev}\sum_sA^{(s,k)}\alpha_i^{(s)}
 &\ge C_{\rm raw}S_{\rm ev}\sum_sA^{(s,k)}t^{(s)}\gamma_i^{(s)}
 \ge \nu h_i.
\end{align*}
Here we used $\alpha_i\ge t\gamma_i$,
$\alpha_i^{(s)}\ge t^{(s)}\gamma_i^{(s)}$,
$C_{\rm raw}t(1+S_{\rm ev})\ge\beta$,
$C_{\rm raw}S_{\rm ev}\ge\nu$, and the definition of $h_i$
in~\eqref{eq:repair-history-representation}. Since
\[
\delta_i=\dfrac Dm-\beta\gamma_i-\nu h_i,
\]
we obtain
\begin{align*}
 &\delta_i+C_{\rm raw}\alpha_i(1+S_{\rm ev})
   +C_{\rm raw}S_{\rm ev}\sum_sA^{(s,k)}\alpha_i^{(s)}\ge\dfrac Dm.
\end{align*}
Consequently, for every fixed deviation $a$,
\[
 d_{i,a}(\widetilde{\vct x})\ge C_{\rm raw}q+\dfrac Dm-e_i^{\rm new}-\sum_se_i^{(s)}.
\]
Taking the minimum over $a$ gives the same bound for
$q_i(\widetilde{\vct x})$. Hence, with
$e_i^{\rm cor}=((C_{\rm raw}q+D/m)-q_i(\widetilde{\vct x}))_+$,
\begin{equation}\label{eq:raw-error-decomposition}
 e_i^{\rm cor}\le e_i^{\rm new}+\sum_se_i^{(s)},
 \qquad E\le\sum_i e_i^{\rm new}+\sum_{s,i}e_i^{(s)}.
\end{equation}
This proves the error decomposition; it yields~\eqref{eq:finite-error-bound} once the estimates in
\eqref{eq:raw-current-total} and \eqref{eq:raw-history-total} are substituted.

\emph{Stage 3: Correction and Retention.}
Apply the forward correction~\eqref{eq:algorithm-forward-correction},
fixing every $e_i^{\rm cor}$ before starting the sweep. Process players
$i=1,\ldots,m$ and mix in a current pure best response with probability
\(\theta_i=\dfrac{e_i^{\rm cor}}{1+\sum_{j\le i}e_j^{\rm cor}}.\)
Before player $i$ is corrected, previous corrections retain at least
$q_i(\widetilde{\vct x})/(1+\sum_{j<i}e_j^{\rm cor})$. Her own
correction raises this to at least
\(\dfrac{q_i(\widetilde{\vct x})+e_i^{\rm cor}} {1+\sum_{j\le i}e_j^{\rm cor}}.\)
Subsequent corrections multiply this bound by retention factors whose
product is $(1+\sum_{j\le i}e_j^{\rm cor})/(1+E)$. This proves
\eqref{eq:corrected-bound}. The sweep also retains the fraction
$1/(1+E)$ of its input labels, giving $\widehat C$.

If $\widehat C\ge c$, apply property~\ref{item:certificate-thinning} of
Lemma~\ref{lem:certificate-coupling}: add independent auxiliary
Bernoulli coins of parameter $c/\widehat C$ to the certificate events,
without changing any strategy coordinate or action leaf. Thus each old
mass becomes $cA^{(s,k)}$, equivalently
$A^{(s,k+1)}=cA^{(s,k)}$, and the new repair event has common mass
$A^{(k,k+1)}=c$. This gives
\eqref{eq:memory-update}
and, finally,
\(\dfrac{C_{\rm raw}\varsigma+D}{1+E} \ge c\varsigma+\dfrac D{1+E} \ge c\varsigma+D-DE.\)
\end{proof}

A failed LP supplies the fresh player law $\vct\lambda$ and conditional
deviations in~\eqref{eq:algorithm-dual-law}. The column bounds
\eqref{eq:finite-fresh-columns} use only this fresh certificate;
Lemma~\ref{lem:history-overlap} then sharpens them to the floor condition
\eqref{eq:finite-floor-feasibility} required by Lemma~\ref{lem:finite-legal-tilt}.
All relevant deficits are nonnegative;~\eqref{eq:dual-columns}
and~\eqref{eq:dual-base} give
\begin{equation}\label{eq:finite-fresh-columns}
 \sum_i\lambda_i d_{ij}(\vct y_j)
 <\dfrac{\varsigma+\mu}{m}=\dfrac{S_u-d}{m}<\dfrac3m,\qquad 
 \lambda_j\le\sum_i\lambda_i d_{ij}(\vct b_j).
\end{equation}
Consequently $\lambda_{\max}<3/m$ before any improved cap is used.
The current base deficit is also below $3/m$. Its bound in~\eqref{eq:dual-base} is
$(\varsigma+\eta\mu)/m$. If $\mu\ge0$, then
$\varsigma+\eta\mu\le\varsigma+\mu=S_u-d<3$; if $\mu<0$,
then $\varsigma+\eta\mu<\varsigma<\sigma<2$. Thus the bound
holds without using positivity of $\mu$, which is proved in
Lemma~\ref{lem:joint-invariant}.

The quantity $r=\max_i\pi_i-1$, where $\pi_i=m\lambda_i$,
measures the concentration of the fresh player law. The interaction
between this law and the stored memory is recorded by
$$
 \tau=\dfrac1m\sum_i\pi_i v_i.
$$
When $H > 0$, $\zeta_i=mh_i/H$ has mean one, and therefore
$$
 \psi=m^{-1}\sum_i\pi_i\zeta_i-1\le r.
$$
At zero memory set $h_i=0$ and $H\psi=0$; the activation rule~\eqref{eq:repair-controls}
requires no division by $H$.

The historical column bound~\eqref{eq:historical-column-certificate}
is maintained by the induction in Lemma~\ref{lem:joint-invariant}.
Assume that every stored repair event has column bounds below $9/m$,
and hence satisfies~\eqref{eq:historical-column-certificate} with $L_{\rm col}=9$.
Lemma~\ref{lem:finite-legal-tilt} will give
$\gamma_i\le3\lambda_i$, transferring~\eqref{eq:finite-fresh-columns}
to the new repair event.
For now, Lemma~\ref{lem:history-overlap}, with
$Y=R_0H$ and $mh_i\le(1+R_0)H$, gives
\[
 (1+r)(1-\vartheta)+H(1+\psi)<S_u-d,
 0\le\vartheta\le\dfrac{(10+R_0)H}{m}<\dfrac{11}{m}.
\]
Expanding the first inequality and using $1+r<3$ gives
\(r+H\psi<C_u(H)-d+(1+r)\vartheta \le C_u(H)-d+\dfrac{33}{m} \le C_u(H)-\dfrac d2,\)
where the last step uses $m\ge66/d$ from~\eqref{eq:finite-player-cutoff}. Thus
$$
 r+H\psi\le C_u(H)-d/2.
$$
Together with the bound $\psi\le r$ above, the identity
\(\dfrac{1+QH}{1+H}(r+H\psi)-(r+QH\psi) =\dfrac{(Q-1)H}{1+H}(r-\psi)\ge0\)
transforms the preceding bound on $r+H\psi$ into the finite floor-feasibility condition
\begin{equation}\label{eq:finite-floor-feasibility}
 r+QH\psi\le R_0.
\end{equation}
This establishes feasibility using only the fresh certificate and the
previously stored repair events.

\begin{lemma}[Legal Tilted Repair]\label{lem:finite-legal-tilt}
Fix a nonterminal state with $m\ge m_0$ and a failed LP whose dual
law $\vct\lambda$ satisfies~\eqref{eq:dual-columns} and
\eqref{eq:finite-floor-feasibility}. Assume the inductive memory bounds
$h_i\ge0$, $H=\sum_i h_i\le L_0$, and
$v_i=mh_i-H\le R_0H$ for every $i$. Define
\begin{equation}\label{eq:finite-tilt}
 Z_0=G_0-Q\tau,
 \chi_i=\pi_i\dfrac{G_0-Qv_i}{Z_0},
 \gamma_i=\dfrac{\chi_i}{m}.
\end{equation}
Then $\vct\gamma$ is a rational probability vector, with
$\gamma_i<3/m$ and $\gamma_i\le3\lambda_i$.  The reverse allocation
$$
 D=\beta(G_0+QH),\qquad
 \delta_i=\dfrac\beta m(G_0-Qv_i-\chi_i)
$$
is nonnegative and has total mass $1-\rho$. The exact memory update~\eqref{eq:memory-update} preserves
the maintained invariant.
\end{lemma}

\begin{proof}
\emph{Normalization and Domination.}
Condition~\eqref{eq:finite-floor-feasibility} gives $Z_0\ge1+r\ge1$.
By the control bounds~\eqref{eq:finite-control-bounds} and the assumed memory invariant,
\(G_0-Qv_i\ge G_0-QR_0L_0>G_0-R_0=1.\)
Moreover $m^{-1}\sum_i\pi_i=1$ and
$m^{-1}\sum_i\pi_i v_i=H\psi$, so averaging the numerator gives
$m^{-1}\sum_i\chi_i=1$.
Since $\pi_i\le1+r\le Z_0$,
\begin{equation}\label{eq:finite-domination}
 \chi_i\le G_0-Qv_i\le G_0+QH<G_0+1<3,
 \dfrac{\chi_i}{\pi_i}\le\dfrac{G_0+QH}{1+r}<3.
\end{equation}
The ratio is used only for $\pi_i>0$; otherwise $\chi_i=0$.
This proves both bounds on $\gamma_i$.  Since the deficits are nonnegative,
the bounds in~\eqref{eq:finite-fresh-columns} transfer to the activated law,
giving a bound of $9/m$. The factor three applies to the fresh certificate
at each generation.

\emph{Allocation and Memory.}
Equation~\eqref{eq:finite-domination} proves $\delta_i\ge0$, while
$\sum_i v_i=0$ and $\sum_i\chi_i=m$ give
$\sum_i\delta_i=\beta R_0=1-\rho$.
Also, the displayed definition of $\delta_i$ is exactly
$\delta_i=D/m-\beta\gamma_i-\nu h_i$, the allocation required by
Lemma~\ref{lem:rawrepair}.

For the coordinate bound after a repair, apply~\eqref{eq:memory-update}:
 $$h_i^+=c(h_i+t\gamma_i)=c\left(h_i+\dfrac{t\chi_i}{m}\right),\qquad 
 H^+=c(H+t),\qquad 
 v_i^+=mh_i^+-H^+=c[v_i+t(\chi_i-1)].
$$
Since $\chi_i\le G_0-Qv_i=1+R_0-Qv_i$ and $1-Q_0>0$,
\begin{align*}
 v_i^+&\le c[(1-Q_0)v_i+tR_0]
 \le cR_0[(1-Q_0)H+t]\le R_0H^+.
\end{align*}
Also $H^+\le c(L_0+t)=L_0$ by~\eqref{eq:finite-potential-controls}.
The descent update~\eqref{eq:algorithm-descent-memory} scales $h_i$, $H$, and $v_i$
equally, preserving the same invariant. At $H=0$,~\eqref{eq:finite-tilt}
gives $\vct\gamma=\vct\lambda$ and $h_i^+=ct\lambda_i$; the first repair gives $H^+=ct>0$.
\end{proof}

The repair has three stages: activation~\eqref{eq:algorithm-activation},
the reverse sweep~\eqref{eq:algorithm-reverse-sweep}, and the forward
correction~\eqref{eq:algorithm-forward-correction}. For a failed LP, first form
\(\vct y_i=(1-t\gamma_i)\vct x_i+t\gamma_i\vct b_i, \qquad P_i=1-\sum_{j\le i}\delta_j,\quad P_0=1.\)
Process players in the order $m,m-1,\ldots,1$, mixing player $i$'s strategy with
her current pure best response with probability $\delta_i/P_{i-1}$.
Let $\widetilde{\vct x}$ be the resulting raw profile. Its correction
deficits from~\eqref{eq:algorithm-correction} are
\[
 C_{\rm raw}=\rho\prod_i(1-t\gamma_i),\qquad 
 e_i=\left(\dfrac{C_{\rm raw}\varsigma+D}{m}-q_i(\widetilde{\vct x})\right)_+,
 \qquad E=\sum_i e_i.
\]
Fix these deficits before the forward sweep. In the order
$1,2,\ldots,m$, mix player $i$'s strategy with her current pure best response
with probability $e_i/(1+\sum_{j\le i}e_j)$.

\begin{lemma}[A Uniform Bound on the Total Deficit]\label{lem:finite-raw-error}
Fix a reachable nonterminal failed-LP state with $m\ge m_0$.
Assume the historical memory invariants and the column-certificate invariant
\eqref{eq:historical-column-certificate}, and use the legal activation
law and reverse allocations of Lemma~\ref{lem:finite-legal-tilt}.
With $L_{\rm col}=9$ and the controls in Subsection~\ref{subsec:proof-controls}, the total deficit
frozen before the forward correction satisfies
\begin{equation}\label{eq:finite-error-bound}
 E\le\dfrac{C_Eu}{m}=\dfrac{120u}{m}.
\end{equation}
For $m\ge1000/u^2$, correction leaves a retained base coefficient of at least
$c$, and auxiliary certificate thinning gives~\eqref{eq:memory-update}, together
with
\begin{equation}\label{eq:finite-sigma-repair}
 \varsigma^+\ge c\varsigma+D-E.
\end{equation}
\end{lemma}

\begin{proof}
\emph{Raw retention and available memory.}
Write $L=L_{\rm col}=9$. Lemma~\ref{lem:finite-legal-tilt} gives
$\gamma_i\le3/m$ and $\gamma_i\le3\lambda_i$.  Since
$t\gamma_i\le3u/m<1/2$ (as $m\ge m_0\ge1000/u^2$),
\begin{equation}\label{eq:raw-odds-bounds}
 \alpha_i:=\dfrac{t\gamma_i}{1-t\gamma_i}\le2t\gamma_i,
 \quad \sum_i\alpha_i\le2t,
 \quad \max_i\alpha_i\le\dfrac{6t}{m}.
\end{equation}
Moreover $\rho\ge1-u\ge3/4$.  Also $QL_0<c<1$ and
$G_0<2$, while $\beta=ct(1+u_0)<(3/5)u(17/16)$; hence
$D=\beta(G_0+QH)<(51/80)u\cdot3<2u$.
Therefore $P_i\ge\rho$ and
$\omega_i=\delta_i/P_i\le(D/m)/\rho<3u/m$, so
\begin{equation}\label{eq:raw-reverse-odds-bounds}
 \omega_i:=\dfrac{\delta_i}{P_i}\le\dfrac{3u}{m},
 \qquad \sum_i\omega_i\le2u,
 \qquad \prod_i(1+\omega_i)=\dfrac1\rho.
\end{equation}
The product identity gives
$\sum_i\omega_i\ge\log(1/\rho)$.  For the upper bound,
$\sum_i\omega_i\le(1-\rho)/\rho\le u/(1-u)<2u$,
using $1-\rho\le u$ and $u\le1/16$.

The controls in~\eqref{eq:finite-retention-controls}
and~\eqref{eq:finite-repair-coefficients} give the two reserves
$$
 \rho e^{-f_0u}-c\ge\dfrac{5u^3}{6},
 \qquad f_0u+\log(1/\rho)\ge u_0+\dfrac{u^2}{2}.
$$
For the first, put $f=f_0$ and $\bar f=1-f$.  Multiplication of the cubic
lower Taylor polynomial for $e^{-fu}$ by $\rho$ gives
\[
 \rho\left(1-fu+\dfrac{f^2u^2}{2}-\dfrac{f^3u^3}{6}\right)-c
 =u^3\left(1-\dfrac f2+\dfrac{f^2}{2}-\dfrac{f^3}{6}
   +\left(\dfrac{\bar f f^3}{6}+\dfrac{\bar f^2f^2}{4}\right)u
   -\dfrac{\bar f^2f^3}{12}u^2\right).
\]
The cubic coefficient is decreasing in $f$ and is at least
$211/250>5/6$ on $[7/12,3/5]$; the last negative term has magnitude at most
$u^2/12\le1/3072<211/250-5/6$, while the preceding higher-order term
is nonnegative.  This proves the first reserve inequality above.
For the second, $-\log\rho\ge1-\rho$ gives
\[
 f_0u+\log(1/\rho)
 \ge u-\dfrac{(1-f_0)^2u^2}{2}
 \ge u-\dfrac{u^2}{2}=u_0+\dfrac{u^2}{2}.
\]  Also $\sum_i\gamma_i^2\le3/m\le L/m$ and
$\log(1-s)\ge-s-s^2$ for $0\le s\le1/2$, so
\begin{equation}\label{eq:raw-retention-lower}
 C_{\rm raw}=\rho\prod_i(1-t\gamma_i)
 \ge c+\dfrac{5u^3}{6}-\dfrac{Lu^2}{m}.
\end{equation}
In particular $C_{\rm raw}\ge c$. By~\eqref{eq:event-memory}
and the second reserve inequality above, the event memory satisfies
$S_{\rm ev}\ge t+\log(1/\rho)>u_0$, so
$C_{\rm raw}t(1+S_{\rm ev})\ge\beta$ and
$C_{\rm raw}S_{\rm ev}\ge\nu$.

\emph{The current-generation error.}
The fresh certificates~\eqref{eq:dual-base} and~\eqref{eq:finite-fresh-columns},
after the domination bound~\eqref{eq:finite-domination}, satisfy
$$
 \sum_i\gamma_i d_i^0\le\dfrac{L}{m},
 \qquad
 \sum_i\gamma_i d_{ij}(\vct b_j)\le\dfrac{L}{m}
 \quad\text{for every }j.
$$
Using~\eqref{eq:raw-odds-bounds},~\eqref{eq:raw-reverse-odds-bounds}, and
$\alpha_i\le2t\gamma_i$, the five terms in
\eqref{eq:raw-current-error} obey, after summing over $i$,
$$ \sum_i\alpha_i d_i^0\le\dfrac{2tL}{m},
 \qquad \sum_i\alpha_i^2\le\dfrac{12t^2}{m},\qquad 
 \sum_i\alpha_i\omega_i
  \le (\max_i\omega_i)\sum_i\alpha_i\le\dfrac{6tu}{m}
  $$
  and 
  $$
 \sum_{i,j\ne i}\alpha_i\alpha_jd_{ij}(\vct b_j)\le\dfrac{4t^2L}{m},\qquad 
 \sum_{i,j\ne i}\alpha_i\omega_jd_{ij}(\vct e_j^{\rm rev})
 \le\dfrac{4tuL}{m}.
$$
Since $t\le u\le1/16$ and $L=9$, their total is at most
\begin{equation}\label{eq:raw-current-total}
 \dfrac{18t+48t^2+42tu}{m}\le\dfrac{24u}{m}.
\end{equation}

The separate factors in~\eqref{eq:raw-current-total} are important.
The fresh certificates above control weighted sums with the current $\gamma_i$,
whereas the raw events carry the odds $\alpha_i$. The inequality
$\alpha_i\le2t\gamma_i$ is the bridge between them; it is applied
once to every activation coordinate. Specifically, the base term obeys
$$
 \sum_i\alpha_i d_i^0
 \le2t\sum_i\gamma_i d_i^0\le\dfrac{2tL}{m}.
$$
Since $\max_i\alpha_i\le6t/m$ and $\sum_i\alpha_i\le2t$,
the same-player activation collision obeys
$$
 \sum_i\alpha_i^2
 \le(\max_i\alpha_i)\sum_i\alpha_i
 \le\dfrac{12t^2}{m}.
$$
Similarly the interaction between an activation and a reverse label
at the same player is at most
$$
 \sum_i\alpha_i\omega_i
 \le(\max_i\omega_i)\sum_i\alpha_i
 \le\dfrac{6tu}{m}.
$$
For distinct players, interchange the $i$ and $j$ sums and use the
fresh column certificate above separately for each fixed $j$. This gives
$$
 \sum_{i,j\ne i}\alpha_i\alpha_jd_{ij}(\vct b_j)
 \le4t^2\sum_j\gamma_j
        \sum_{i\ne j}\gamma_i d_{ij}(\vct b_j)
 \le\dfrac{4t^2L}{m}.
$$
The reverse-change odds satisfy $\sum_j\omega_j\le2u$
by~\eqref{eq:raw-reverse-odds-bounds}.
Since~\eqref{eq:finite-fresh-columns}, transferred by~\eqref{eq:finite-domination},
is valid for \emph{every} opponent action, it
holds in particular for the fixed reverse best-response action at $j$.
Thus
$$
 \sum_{i,j\ne i}\alpha_i\omega_jd_{ij}(\vct e_j^{\rm rev})
 \le2t\sum_j\omega_j\sum_{i\ne j}\gamma_i
                 d_{ij}(\vct e_j^{\rm rev})
 \le\dfrac{4tuL}{m}.
$$
Adding the five bounds above gives exactly
$[2Lt+(12+4L)t^2+(6+4L)tu]/m$, which is the coefficient
in~\eqref{eq:raw-current-total} after setting $L=9$.
Because $t\le u\le1/16$, the last coefficient is bounded by
$[18+90/16]u<24u$.

\emph{The historical error.}
Summing over historical generations $s$, write
$H_i=\sum_{s\in\mathcal R^{(k)}} A^{(s,k)}\alpha_i^{(s)}$.  These aggregate odds satisfy
$\sum_iH_i\le2H$ by~\eqref{eq:repair-history-representation}
and the odds bound~\eqref{eq:raw-odds-bounds} at each stored generation.
The stored certificates~\eqref{eq:historical-column-certificate} give
$$
 \sum_{i\ne j}\alpha_i^{(s)}d_{ij}^{(s)}(\vct b_j)\le\dfrac{2t^{(s)} L}{m}
 \quad\text{for every }s,j.
$$
The four terms in~\eqref{eq:raw-history-error}, summed over $i$ and $s$,
are bounded respectively by
\begin{align*}
 \sum_{s,i}A^{(s,k)}\alpha_i^{(s)}\alpha_i&\le\dfrac{12tH}{m},
 &\sum_{s,i}A^{(s,k)}\alpha_i^{(s)}\omega_i&\le\dfrac{6uH}{m},\\
 \sum_{s,i,j\ne i}A^{(s,k)}\alpha_i^{(s)}\alpha_jd_{ij}^{(s)}(\vct b_j)
 &\le\dfrac{4tLH}{m},
 &\sum_{s,i,j\ne i}A^{(s,k)}\alpha_i^{(s)}\omega_jd_{ij}^{(s)}(\vct e_j^{\rm rev})
 &\le\dfrac{4uLH}{m}.
\end{align*}
Thus the historical contribution is at most
\begin{equation}\label{eq:raw-history-total}
 \dfrac{(48t+42u)H}{m}\le\dfrac{90uH}{m}.
\end{equation}

For clarity,~\eqref{eq:raw-history-total} retains the full sum over old generations
until after~\eqref{eq:historical-column-certificate} has been applied.
By this certificate and $\alpha_i^{(s)}\le2t^{(s)}\gamma_i^{(s)}$,
for each old generation $s$ and fixed opponent $j$ we have
$$
 \sum_{i\ne j}\alpha_i^{(s)}d_{ij}^{(s)}(\vct y_j)
 \le\dfrac{2t^{(s)}L}{m}.
$$
This statement is independent of the current activation law and
holds even when $\vct y_j$ is a reverse best response selected only
at the current iteration. The old column is evaluated at generation
$s$, and property~\ref{item:certificate-override} of Lemma~\ref{lem:certificate-coupling}
justifies taking this override without
changing the other old conditional leaves.

Summing the retained odds gives
$\sum_{s,i}A^{(s,k)}\alpha_i^{(s)}\le2H$.
Hence the two errors in which the current changed label is at the
same player as the old activation are bounded by
$$
 \sum_{s,i}A^{(s,k)}\alpha_i^{(s)}(\alpha_i+\omega_i)
 \le\dfrac{(12t+6u)H}{m}.
$$
The opponent-activation part is at most
$$
 \sum_{s,i,j\ne i}A^{(s,k)}\alpha_i^{(s)}\alpha_j
 d_{ij}^{(s)}(\vct b_j)
 \le\dfrac{4tL}{m}\sum_s A^{(s,k)}t^{(s)}
 =\dfrac{4tLH}{m}.
$$
The reverse-label part is similarly at most
$$
 \sum_{s,i,j\ne i}A^{(s,k)}\alpha_i^{(s)}\omega_j
 d_{ij}^{(s)}(\vct e_j^{\rm rev})
 \le\dfrac{2L}{m}\sum_s A^{(s,k)}t^{(s)}\sum_j\omega_j
 \le\dfrac{4uLH}{m}.
$$
The three bounds above sum to
$[(12+4L)t+(6+4L)u]H/m=(48t+42u)H/m$.
Thus the historical contribution depends only on the single scalar
$H=\sum_sA^{(s,k)}t^{(s)}$, keeping the total correction error bounded
across successive repairs.

Because $H\le L_0<1$, substituting~\eqref{eq:raw-current-total} and
\eqref{eq:raw-history-total} into~\eqref{eq:raw-error-decomposition} gives
$E\le114u/m\le120u/m$, proving~\eqref{eq:finite-error-bound}.

\emph{Retention after correction.}
By~\eqref{eq:corrected-bound}, the correction retains a coefficient of
$C_{\rm raw}/(1+E)$. From~\eqref{eq:raw-retention-lower},
\eqref{eq:finite-error-bound}, and $c<1$,
$$
 C_{\rm raw}-c(1+E)
 \ge\dfrac{5u^3}{6}-\dfrac{Lu^2+C_Ecu}{m}
 \ge u^3\left(\dfrac56-\dfrac{Lu+C_Ec}{1000}\right)>0,
$$
where $m\ge1000/u^2$ by~\eqref{eq:finite-player-cutoff} and $u\le1/16$. Therefore the corrected
coefficient is at least $c$.

To see that the final strict inequality above holds under the
stated numerical controls, use $c<1$ and $u\le1/16$ to get
$$
 \dfrac56-\dfrac{Lu+C_Ec}{1000}
 >\dfrac56-\dfrac{9/16+120}{1000}>0.
$$
The lower bound on the player count thus compensates for the
finite-dimensional interaction errors. Since $\widehat C\ge c$,
property~\ref{item:certificate-thinning} of Lemma~\ref{lem:certificate-coupling}
thins the auxiliary certificates to record a retention factor of exactly $c$.
The correction bound~\eqref{eq:corrected-bound} and $D\le2u\le1$ give
\(\varsigma^+\ge\dfrac{C_{\rm raw}\varsigma+D}{1+E} \ge c\varsigma+D-E.\)
The certificate thinning therefore gives the exact update~\eqref{eq:memory-update}.
\end{proof}

\begin{proof}[Proof of Proposition~\ref{prop:repair}]
The failed-LP dual columns~\eqref{eq:dual-columns} are established in Lemma~\ref{lem:descent}.
The historical overlap~\eqref{eq:historical-overlap} and floor estimate
\eqref{eq:finite-floor-feasibility} yield the legal tilted law;
Lemma~\ref{lem:finite-legal-tilt} proves nonnegativity of the reverse
allocation and preserves the invariant.  Lemma~\ref{lem:finite-raw-error}
then bounds the current and historical correction deficits by $120u/m$ and
combines the two sweeps with~\eqref{eq:memory-update}, giving the
satisfaction inequality~\eqref{eq:finite-sigma-repair} and hence~\eqref{eq:main-repair}.
\end{proof}

\subsection{Scalar Certificate and Potential Function}\label{subsec:potential-termination}

With the local descent and repair estimates in place, we now construct
a potential function that certifies progress in both branches.
We first determine the scalar approximation coefficient in
Theorem~\ref{thm:squares}, then pass to finite steps by controlling the error
uniformly down to zero memory in Lemma~\ref{lem:finite-uniform}.
This yields the potential progress bounds of Proposition~\ref{prop:potential}.

The historical state measures how much useful memory from previous
repairs remains in the current profile. Let $k$ denote the outer
algorithm iteration, and let $\mathcal R^{(k)}$ be the set of repair
generations created before the current state at iteration $k$. The
repair-event index is denoted by $s$ throughout. For each $s\in\mathcal R^{(k)}$,
the repair event has an activation law $\vct\gamma^{(s)}$, an intensity
$t^{(s)}$, and a retained common coefficient $A^{(s,k)}$. Define
$$
 \begin{aligned}
 h_i^{(k)}&=\sum_{s\in\mathcal R^{(k)}}A^{(s,k)}t^{(s)}\gamma_i^{(s)},
 &H^{(k)}&=\sum_ih_i^{(k)}=\sum_{s\in\mathcal R^{(k)}}A^{(s,k)}t^{(s)},\\
 v_i^{(k)}&=mh_i^{(k)}-H^{(k)},
 &\sum_i v_i^{(k)}&=0.
 \end{aligned}
$$
In the remainder of a local calculation, the current iteration $k$ is
fixed and we suppress the superscript $(k)$, writing
$A^{(s,k)},h_i,H,v_i$ for $A^{(s,k)},h_i^{(k)},H^{(k)},v_i^{(k)}$;
all sums over $s$ are then over $\mathcal R^{(k)}$. When we compare two
successive states, we retain the superscript $(k)$ explicitly.
The retained-event construction is given in Lemma~\ref{lem:certificate-coupling},
and the column certificates are stated in~\eqref{eq:historical-column-certificate}.

A failed descent LP supplies, by Lemma~\ref{lem:descent}, a player law $\vct\lambda$ and conditional
deviations. Put
\[\pi_i=m\lambda_i, \qquad r=\max_i\pi_i-1, \qquad \tau =\dfrac1m\sum_i\pi_i v_i.\]
When $H >0 $, $\zeta_i=mh_i/H$ has uniform-player mean one, as
does $\pi_i$. Since $\pi_i\le1+r$ and $\zeta_i\ge0$,
$$
 \tau / H =\dfrac1m\sum_i\pi_i\zeta_i-1\le r.
$$

The excess $r$ measures the concentration of the dual law, while $\tau $
measures its overlap with the historical information. The inequality above
relates these two quantities.

In the large-player limit with vanishing repair steps, the historical
overlap bound takes the form
$$
 r+\tau \le C(H),
 \qquad S(H)=1+H+C(H).
$$
Lemma~\ref{lem:history-overlap} supplies the finite version~\eqref{eq:historical-overlap}, and
the controls in Subsection~\ref{subsec:proof-controls} reserve strict slack for its $1/m$ error.

Fix $0<f<1$ and set $R=(1-f)/f$. In the scalar model, the tilted
activation is
$$
 \gamma_i=\lambda_i\dfrac{1-v_i}{1-\tau }.
$$
The memory cap $v_i\le RH$, together with $0\le H\le f$, makes each
numerator positive. The common reverse allocation is feasible if
$$
 f(1+r)\le1-\tau ,
 \quad\text{equivalently}\quad
 r+\dfrac{\tau }{f}\le R.
$$
This condition also makes the normalization positive.

\begin{lemma}[The Maximal Scalar Threshold]
\label{lem:threshold}
The moment constraints $\tau\le rH$ and $r+\tau\le C(H)$
imply the reverse-allocation condition $r+\tau/f\le R$ when
\begin{equation}
 C_f(H)=\dfrac{(1-f)(1+H)}{f+H},
 \qquad S_f(H)=\dfrac{(1+H)^2}{f+H}.
 \label{eq:new-threshold}
\end{equation}
This is the largest threshold implied by those two moment constraints.
\end{lemma}
\begin{proof}
Since $\tau \le rH$,
\[r+\dfrac{\tau}{f} \le\dfrac{1+H/f}{1+H}(r+\tau ) \le\dfrac{1+H/f}{1+H}C_f(H)=R.\]
The difference in the first inequality is
$(1/f-1)(rH-\tau )/(1+H)\ge0$.
For sharpness when $H>0$ within the recorded constraints, take
$r=\tau/H=C/(1+H)$. The mean-one law
$\pi=\zeta\in\{0,1+r\}$, with probability $1/(1+r)$ on its positive
value, realizes these moments. Increasing $C$ beyond $C_f(H)$ then
violates the reverse-allocation condition. When $H=0$, the moment constraint
forces $\tau=0$; taking $r=C>R=C_f(0)$ already violates the same
floor condition. One may realize this scalar boundary case with
zero memory and a mean-one player law supported on $0$ and $1+r$.
This establishes sharpness in the scalar relaxation.
\end{proof}

The drift relations~\eqref{eq:analysis-descent-ode} and~\eqref{eq:analysis-repair-ode}
describe how descent and repair change satisfaction and history. The
potential defined below accounts for both effects through the threshold
$S_f$ in~\eqref{eq:new-threshold}. Its key property is the derivative
relation with $S_f$ derived below.

For $0<H\le f$, define
$$
 \Phi_f(\varsigma,H)=
 \dfrac{\varsigma-1/f}{H}
 +\dfrac{2f-1}{f^2}\log(H/f)
 +\dfrac{(1-f)^2}{f^2}\log(1+H/f).
$$
Writing $F_f=\Phi_f-\varsigma/H$, we obtain the partial-fraction decomposition
\[\dfrac{S_f(H)}{H^2} =\dfrac1{fH^2}+\dfrac{2f-1}{f^2H} +\dfrac{(1-f)^2}{f^2(H+f)}.\]
Consequently
$
 H^2F_f'(H)=S_f(H), \partial_H\Phi_f(\varsigma,H)=\dfrac{S_f(H)-\varsigma}{H^2}.
$

The limiting repair directions from~\eqref{eq:analysis-repair-ode} are
\(\dot H=f-H, \dot\varsigma\ge1+H-\varsigma.\)
The first equation balances newly created memory against the loss of
old memory. Substitution into the displayed formula for $\partial_H\Phi_f$ gives
\(\dot\Phi_f \ge\dfrac{1+H-\varsigma}{H} +\dfrac{(f-H)[S_f(H)-\varsigma]}{H^2}.\)
Collecting the terms in $\varsigma$ yields
\begin{equation}
 \dot\Phi_f\ge\dfrac f{H^2}\bigl[K_f(H)-\varsigma\bigr],\qquad 
 K_f(H)=\dfrac{(1+H)[f+(2f-1)H]}{f(f+H)}.
 \label{eq:continuous-progress}
\end{equation}
Equivalently,
\(K_f(H)=\dfrac Hf(1+H)+\left(1-\dfrac Hf\right)S_f(H) =S_f(H)-\dfrac HfC_f(H).\)
Thus a target below $\min_HK_f(H)$ has strictly positive repair
progress. The finite-step counterpart is~\eqref{eq:finite-repair-identity}.

It remains to choose $f$ so that the smallest value of $K_f$ on
$[0,f]$ is as large as possible, as specified in~\eqref{eq:maxmin}.
The function~\eqref{eq:one-variable} identifies the candidate optimum;
the square identities of Theorem~\ref{thm:squares} prove its global optimality.
Define
\begin{equation}
 H_{\rm c}(s)=\dfrac{(1+s)(1-2s^2)}{1-s},
 \qquad 0\le s\le\dfrac12.
 \label{eq:one-variable}
\end{equation}
Its derivative can be expressed through a cubic polynomial:
\[
 H_{\rm c}'(s)=\dfrac{2Q_{\rm c}(s)}{(1-s)^2},
 Q_{\rm c}(s)=2s^3-2s^2-2s+1,
 Q_{\rm c}'(s)=2(3s+1)(s-1).
\]
The polynomial $Q_{\rm c}$ strictly decreases on $[0,1/2]$ and has
opposite signs at the endpoints. Its unique zero $s_{\star}$ is therefore
the unique maximizer of $H_{\rm c}$. Put $f_{\star}=1-s_{\star}$.
The signs at $2/5$ and $5/12$ give
$$
 \dfrac25<s_{\star}<\dfrac5{12},
 \qquad \dfrac7{12}<f_{\star}<\dfrac35,\qquad 
 2f_{\star}^3-4f_{\star}^2+1=0.
$$
Reduction by the cubic above gives
$$
 H_{\rm c}(s_{\star})
 =6s_{\star}^2+4s_{\star}-1
   =2s_{\star}(1+s_{\star})^2
 =9-16f_{\star}+6f_{\star}^2
   =\dfrac{(2-f_{\star})(1-f_{\star})}{f_{\star}^2}.
$$
Denote this value by $\sigma_{\star}$.

\begin{samepage}
\begin{theorem}[Optimality of the Scalar Certificate]
\label{thm:squares}
For every $H\ge0$,
\begin{equation}
 K_{f_{\star}}(H)-\sigma_{\star}
 =\dfrac{(2f_{\star}-1)(H-s_{\star})^2}{f_{\star}(f_{\star}+H)}.
 \label{eq:primal-square}
\end{equation}
For every $f>0$,
\begin{equation}
 \sigma_{\star}-K_f(s_{\star})
 =\dfrac{\sigma_{\star}(f-f_{\star})^2}{f(f+s_{\star})}.
 \label{eq:dual-square}
\end{equation}
It follows that
\begin{equation}
 \sigma_{\star}
 =\max_{0<f<1}\;\min_{0\le H\le f}K_f(H)=\max_{0\le s\le1/2}H_{\rm c}(s),
 \label{eq:maxmin}
\end{equation}
and the maximizing activation parameter is uniquely $f_{\star}$.
\end{theorem}
\end{samepage}
\begin{proof}
Multiplying~\eqref{eq:primal-square} and~\eqref{eq:dual-square} by their
positive denominators reduces each assertion to a polynomial identity.
Substitute $s_{\star}=1-f_{\star}$ and
$\sigma_{\star}=9-16f_{\star}+6f_{\star}^2$; the differences are divisible by
$2f_{\star}^3-4f_{\star}^2+1$.

Since $2f_{\star}-1>0$ and $0<s_{\star}<f_{\star}$, identity~\eqref{eq:primal-square} proves
the lower bound in~\eqref{eq:maxmin}, with the unique minimizing state
$H=s_{\star}$. If $f\ge s_{\star}$, identity~\eqref{eq:dual-square} proves
$\min_{0\le H\le f}K_f(H)\le K_f(s_{\star})\le\sigma_{\star}$.
If $f<s_{\star}<1/2$, then
$K_f(f)=1+f<3/2<H_{\rm c}(2/5)<\sigma_{\star}$.
For the second equality in~\eqref{eq:maxmin}, use
Lemma~\ref{lem:fixed-activation}. On the middle branch
$f\in[1/\sqrt3,2/3]$, set
$s=\dfrac12\sqrt{4-2/f}$; then the middle-branch value in~\eqref{eq:fixed-activation-value} is
$H_{\rm c}(s)$. The range of this substitution is
$[\dfrac12\sqrt{4-2\sqrt3},1/2]$, which contains $s_{\star}$;
$H_{\rm c}$ is increasing up to $s_{\star}$, so the omitted interval
cannot contain a larger value. On the two outer branches the values are
$1+f\le1+1/\sqrt3<\sigma_{\star}$ and
$1/f\le3/2<\sigma_{\star}$, respectively. Hence
$\max_{0<f<1}U(f)=\max_{0\le s\le1/2}H_{\rm c}(s)$,
and equality in the upper bound requires $f=f_{\star}$.
\end{proof}

For $k=9-16f+6f^2$, the polynomial identity
$$ 
 k^3-19k^2+27k+1
 =(2f^3-4f^2+1)
 (108f^3-648f^2+1152f-566)
$$ 
proves~the defining equation of Theorem~\ref{thm:main}.
The lower bound
$\sigma_{\star}>H_{\rm c}(2/5)=119/75>1$ follows from
$H_{\rm c}'(2/5)>0$.
At $H=f_{\star}$, the square in~\eqref{eq:primal-square} is strictly
positive, so $\sigma_{\star}<K_{f_{\star}}(f_{\star})=1+f_{\star}<2$.
For $p(k)=k^3-19k^2+27k+1$, one has $p(1)=10>0$ and $p(2)=-13<0$.
Moreover, $p'(1)=-8$ and $p''(k)=6k-38<0$ on $[1,2]$, so $p$ is
strictly decreasing throughout that interval. Thus $\sigma_{\star}$ is the
unique root in $(1,2)$, with $\sigma_{\star}\approx1.5867$.

\begin{lemma}[The Value for a Fixed Activation Parameter]
\label{lem:fixed-activation}
The scalar value $U(f)=\min_{0\le H\le f}K_f(H)$ is
\begin{equation}
 U(f)=
 \begin{cases}
 1+f,&0<f\le\tfrac1{\sqrt3},\\[5pt]
 5-4f-\dfrac1f+2(1-f)\sqrt{4-\dfrac2f},
     &\tfrac1{\sqrt3}\le f\le\tfrac23,\\[7pt]
 \dfrac1f,&\tfrac23\le f<1.
 \end{cases}
 \label{eq:fixed-activation-value}
\end{equation}
\end{lemma}
\begin{proof}
With $x=H/f$,
\(K_f(fx)=\dfrac1f+(2f-1)x-\dfrac{2(1-f)^2x}{f(1+x)}.\)
Its first two derivatives are
\[
 \dfrac{\dd}{\dd x}K_f(fx)
 =2f-1-\dfrac{2(1-f)^2}{f(1+x)^2},\qquad 
 \dfrac{\dd^2}{\dd x^2}K_f(fx)
 =\dfrac{4(1-f)^2}{f(1+x)^3}>0.
\]
Checking the derivative at $x=0,1$ gives the three cases
in~\eqref{eq:fixed-activation-value}. In the interior case, put $z=\sqrt{4-2/f}$.
Then $U(f)=(2-z^2)(2+z)/[2(2-z)]$, which becomes
$H_{\rm c}(s)$ from~\eqref{eq:one-variable} under $z=2s$.
\end{proof}

The threshold~\eqref{eq:new-threshold} has the strict reserve used for
the activation choice~\eqref{eq:finite-activation-choice}:
$$
 \min_{0\le H\le f_{\star}}S_{f_{\star}}(H)
 =\dfrac1{f_{\star}}>\sigma_{\star},
 \dfrac1{f_{\star}}-\sigma_{\star}
 =\dfrac{(2f_{\star}-1)s_{\star}^2}{f_{\star}^2}.
$$
Indeed,
$S_f'(H)=(1+H)(2f+H-1)/(f+H)^2>0$ for $f=f_{\star}$,
and the gap follows from~\eqref{eq:primal-square} at zero.

\begin{remark}[Scope of the Matching Bounds]
\label{rem:scope}
Identities~\eqref{eq:primal-square} and~\eqref{eq:dual-square} solve the max--min problem
in~\eqref{eq:maxmin}. The state $s_{\star}$ therefore certifies optimality
within the scalar certificate~\eqref{eq:continuous-progress}.
\end{remark}

To prove Proposition~\ref{prop:potential}, we use the controls
in~\eqref{eq:finite-reweighting-controls} and~\eqref{eq:finite-potential-controls}
to define the finite-step potential for $H>0$:
\begin{equation}\label{eq:finite-potential-definition}
 F_u(H)=-\dfrac{G_0}{H}
       +(1-a_u)\log(H/f_0)+a_u\log(1+QH),\qquad 
 \Phi_u(\varsigma,H)=\dfrac{\varsigma}{H}+F_u(H).
\end{equation}
Its derivative is tied exactly to the descent threshold~\eqref{eq:algorithm-threshold}:
\begin{equation}\label{eq:finite-synchronization}
 H^2F_u'(H)=S_u(H).
\end{equation}
For the ideal part $\varsigma^+=c\varsigma+D$ of~\eqref{eq:finite-sigma-repair},
with $D$ defined in~\eqref{eq:repair-controls}, write
$T=c(H+t)$ and $M=H(H+t)/t$. The coefficient governing its potential
increment is
\begin{equation}\label{eq:finite-repair-coefficient}
 K_u(H)=G_0+\dfrac{H[D-(1-c)G_0]}{ct}
 +M\left[(1-a_u)\log\dfrac TH
       +a_u\log\dfrac{1+QT}{1+QH}\right].
\end{equation}
Direct algebra yields
\begin{equation}\label{eq:finite-repair-identity}
 \Phi_u^+-\Phi_u
 =\dfrac{t}{H(H+t)}[K_u(H)-\varsigma].
\end{equation}
The actual correction incurs the loss $E$ in
\eqref{eq:finite-sigma-repair}. It decreases the coefficient in
\eqref{eq:finite-repair-identity} by at most
\begin{equation}\label{eq:finite-Ku-error}
 \dfrac{HE}{ct}\le\dfrac{C_EH}{cf_0m}
 \le\dfrac{128}{m}\le d,
\end{equation}
using $H\le L_0<f_0$, $c\ge15/16$, and the cutoff
$m\ge400/d$ from~\eqref{eq:finite-player-cutoff}.

\begin{lemma}[Uniform Control Down to Zero Memory]\label{lem:finite-uniform}
For $f_0\in[7/12,3/5]$, $0<u\le1/16$, and $0\le H\le L_0$,
\begin{equation}\label{eq:finite-uniform-Ku}
 \left|K_u(H)-K_{f_0}(H)\right|\le32u,
\end{equation}
where $K_u(0)=G_0$ denotes its continuous extension.
\end{lemma}

\begin{proof}
Put \(f=f_0\), \(R=(1-f)/f\), \(Q_\star=1/f\), and
\(a_\star=R^2\).  All estimates below hold for
\(0<u\le1/16\) and \(7/12\le f\le3/5\); in particular
\(c>15/16\), \(f\le3/5\), and \(1+u-u^2>1\).

First,
\[
 c(1+u_0)=1-\dfrac32u^2+\dfrac12u^3-\dfrac32u^4+u^5
 \quad\text{and}\quad
 c(1+u_0)\ge1-\dfrac32u^2 .
\]
Since
\[
 R_0
 =R\,\dfrac{1-(1-f)u/2}{c(1+u_0)},
\]
the last display, \(1-(1-f)u/2\ge1-u/2\), and
\(1/(1-\dfrac32u^2)\le1+2u^2\) imply
\begin{equation}\label{eq:finite-parameter-approximation}
 |R_0-R|\le u,\qquad 0<R_0<1,\qquad G_0=1+R_0<2 .
\end{equation}
Furthermore
\[
 Q=\dfrac{1-u}{f(1+u-u^2)},\qquad
 |Q-Q_\star|
 =\dfrac{2u-u^2}{f(1+u-u^2)}
 \le \dfrac{2u}{f}<4u .
\]
Because \(0<R\le 5/7\), \(Q_\star-1=R\), and \(Q-1<1\), it follows that
\[
 |a_u-a_\star|
 \le |R_0-R|\,|Q-1|+R\,|Q-Q_\star|
 \le5u .
\]
Writing \(b=(1-c)/u=1-u/2+u^2\), we have
\[
 b-c=\dfrac u2+\dfrac{u^2}2+u^3,\qquad
 L_0=\dfrac{cf}{b},\qquad
 0\le f-L_0=\dfrac{f(b-c)}b\le u,
\]
and
\[
 \left|\dfrac{1-c}{ct}-\dfrac1f\right|
 =\dfrac{b-c}{cf}\le2u .
\]
Moreover $f(1+u-u^2)<(3/5)(17/16)<1-u$, so $Q>1$,
while $Q<1/(7/12)<2$; together with $R_0<1$ this gives
$0<a_u<1$.  Finally $b-c=u/2+u^2/2+u^3>0$, so
$L_0=cf/b<f$ exactly.

For \(0<H\le L_0\), set \(T=c(H+t)\) and
\(M=H(H+t)/t\).  Since
$$
 T-H=ct\left(1-\dfrac H{L_0}\right),\qquad H\le T\le L_0 ,
$$
the integral bounds for \(1/s\) give
\[
 H\left(1-\dfrac H{L_0}\right)
 \le M\log\dfrac TH
 \le T\left(1-\dfrac H{L_0}\right).
\]
The interval between the two endpoints has length at most
\(ct\le u\), and its lower endpoint differs from
\(G_{1,\star}=H(1-H/f)\) by at most
\(H^2(f-L_0)/(fL_0)\le u\).  Consequently
\[
 |G_1-G_{1,\star}|\le2u,\qquad
 G_1=M\log(T/H).
\]
For the other logarithm, integration of $Q/(1+Qs)$ with the same expression for $T-H$ yields
\[
 \dfrac{QHT}{1+QT}\left(1-\dfrac H{L_0}\right)
 \le G_2\le
 \dfrac{QHT}{1+QH}\left(1-\dfrac H{L_0}\right),
\]
where \(G_2=M\log((1+QT)/(1+QH))\).  The interval has length at most
\(2u\): use \(Q<2\), \(H,T<3/5\), and \(T-H\le u\).  For
\(\Gamma(Q,T,\ell)=QHT\ell/(1+QT)\), on this box
\[
 |\partial_Q\Gamma|\le1,\qquad
 |\partial_T\Gamma|\le2,\qquad
 |\partial_\ell\Gamma|\le1 .
\]
The changes \(|Q-Q_\star|\), \(|T-H|\), and
\[
 \left|\left(1-\dfrac H{L_0}\right)-\left(1-\dfrac Hf\right)\right|
 =\dfrac{H(f-L_0)}{fL_0}\le \dfrac{f-L_0}{f}
 =\dfrac{b-c}{b}\le u
\]
(the last inequality follows from
$ub-(b-c)=u(1/2-u)\ge0$) are at most \(4u,u,u\), respectively.  Hence the lower endpoint differs
from
\(G_{2,\star}=Q_\star H^2(1-H/f)/(1+Q_\star H)\) by at most \(7u\),
and
\[
 |G_2-G_{2,\star}|\le10u.
\]
All these estimates extend to \(H=0\), since
\(H(H+t)\log(T/H)/t\to0\) and the analogous second product tends to zero.

Finally, the algebraic term in $K_u$ from~\eqref{eq:finite-repair-coefficient} is
\[
 \dfrac{H[D-(1-c)G_0]}{ct}
 =HG_0\left(1+u_0-\dfrac{1-c}{ct}\right)
   +\dfrac{1-u}{f}H^2 .
\]
Put \(Q_\star=1/f\) and \(r_c=(1-c)/(ct)\), and define
\[
 A_\star(H)=H Q_\star(1-Q_\star)+\dfrac{H^2}{f}.
\]
The limiting algebraic term is \(A_\star(H)\).  Since
\(|G_0-Q_\star|\le u\), \(|r_c-Q_\star|\le2u\), \(u_0\le u\),
\(H\le3/5\), \(Q_\star\le12/7\), and \(R\le5/7\),
\[
\left|1+u_0-r_c\right|\le R+3u<1.
\]
Consequently,
$$
\left|\dfrac{H[D-(1-c)G_0]}{ct}-A_\star(H)\right|
\le H\left[u+3Q_\star u\right]+\dfrac{H^2u}{f}
\le \left(\dfrac{753}{175}\right)u<5u.
$$
The leading constant differs from \(1/f\) by at most \(u\).  The two
logarithmic terms contribute at most \(2u+10u\), and changing
\(a_\star\) to \(a_u\) contributes at most
\[
 |a_u-a_\star|\,|G_{2,\star}-G_{1,\star}|\le10u,
\]
because both limiting products have absolute value at most one.  The
total is at most \(28u\), hence the asserted \(32u\) bound.  Substituting
the limiting terms gives $K_f(H)$
from~\eqref{eq:continuous-progress}.
\end{proof}

Because $u\le d/100$, the uniform estimate~\eqref{eq:finite-uniform-Ku}
and the activation choice~\eqref{eq:finite-activation-choice} give, for every
$0\le H\le L_0$,
\[
 K_u(H)\ge K_{f_0}(H)-32u
 >\sigma+7d-0.32d>\sigma+6d.
\]
The bound $|R_0-(1-f_0)/f_0|\le u$ in~\eqref{eq:finite-parameter-approximation} gives
$G_0>1/f_0-u>\sigma+7d-0.01d>\sigma+6d$.  Hence
\begin{equation}\label{eq:finite-strict-reserves}
 K_u(H)>\sigma+6d,\qquad G_0>\sigma+6d.
\end{equation}
In particular every nonterminal state has
$\mu\ge G_0-\sigma-d>5d>0$.  This also verifies the positivity
hypotheses used in Lemmas~\ref{lem:descent} and~\ref{lem:finite-raw-error}.

Add a reciprocal term to~\eqref{eq:finite-potential-definition}:
\begin{equation}\label{eq:finite-tilted-potential}
 \widehat\Phi_u=\Phi_u+2d/H.
\end{equation}

\begin{proof}[Proof of Proposition~\ref{prop:potential}]
At a nonterminal state $\varsigma<\sigma$, monotonicity
in~\eqref{eq:finite-threshold-bounds}, the reserve~\eqref{eq:finite-strict-reserves},
and $S_u(0)=G_0$ give
\[
 \mu=S_u(H)-\varsigma-d\ge G_0-\sigma-d>5d.
\]
For a passing LP put \(Z=\sum_jZ_j\) and
\(P=\prod_j(1-\eta Z_j)\).  Since \(P\ge1-\eta Z\) and
\(\eta Z\le1/4\),
\[
 e:=P^{-1}-1\le\dfrac{\eta Z}{1-\eta Z}
 \le\eta Z+2\eta^2 .
\]
Using $H^+=PH$ from~\eqref{eq:algorithm-descent-memory}, together with the synchronization
identity~\eqref{eq:finite-synchronization} and the monotonicity of $S_u$, we obtain
\[
 H(\widehat\Phi_u^+-\widehat\Phi_u)
 \ge \eta(Z\varsigma+\mu)-(S_u(H)-2d)e .
\]
Substitute \(S_u(H)=\varsigma+\mu+d\) and use \(S_u(H)<3\) together with the last
bound on \(e\):
\[
 H(\widehat\Phi_u^+-\widehat\Phi_u)
 \ge\eta\bigl[(1-Z)\mu+Zd\bigr]-6\eta^2
 \ge\dfrac{\eta d}{2},
\]
where \(\eta\le d/24\) and \(\mu>5d\) were used in the last step.  Here \(S_u(H)<3\) follows from \(Q>1\), \(R_0<1\), and \(H\le L_0<f_0\le3/5\): indeed \(S_u(H)\le1+H+R_0<13/5\).

For a failed LP, the ideal repair identity~\eqref{eq:finite-repair-identity}
and correction estimate~\eqref{eq:finite-Ku-error} give a loss of at most
$d$ in the coefficient $K_u(H)$. The reciprocal
term contributes
\[
 \dfrac{2dt}{H+t}\left(\dfrac{(1-c)H}{ct}-1\right)\ge
 -\dfrac{2dt}{H+t},
\]
because \(H\le L_0=ct/(1-c)\).  Since
\(K_u(H)>\sigma+6d\) and \(\varsigma<\sigma\), the resulting coefficient
is at least \(3d\), and
\[
 H(\widehat\Phi_u^+-\widehat\Phi_u)
 \ge\dfrac{t}{H+t}\,3d
 \ge3d(1-c)>0 .
\]
This proves both asserted inequalities.
\end{proof}

\subsection{Joint Induction for Reachable States}
\label{subsec:closed-induction}

We now show that the hypotheses of the preceding local estimates hold
along every run. Since the repair estimates use old certificates and
each new certificate is obtained from a legal repair, we establish
both properties together by joint induction in
Lemma~\ref{lem:joint-invariant}. Throughout, assume $m\ge m_0$ and use
the rational controls of Subsection~\ref{subsec:proof-controls}.

For each iteration $k$, let $\mathcal R^{(k)}$ consist precisely of the
indices $s<k$ at which a repair was executed. When such a repair is
performed, its data $\vct x^{(s)}$, $(\vct b_i^{(s)})_i$,
$\vct\gamma^{(s)}$ and $t^{(s)}=t$ are conceptually frozen. The
analysis also retains their common masses $A^{(s,k)}$. Starting from
$A^{(s,s+1)}=c$, these masses follow the scalar retention updates
in~\eqref{eq:algorithm-descent-memory} and~\eqref{eq:joint-new-masses}.
The action law is always the current product profile $\vct x^{(k)}$.

\begin{lemma}[Joint Reachable-State Invariant]
\label{lem:joint-invariant}
At the beginning of every iteration $k$ reached by
Algorithm~\ref{alg:main} with $m\ge m_0$, the following assertions
hold simultaneously.
\begin{enumerate}[label=\textup{(I\arabic*)},leftmargin=2.7em]
\item The profile $\vct x^{(k)}$ is a rational product profile, all
stored masses are nonnegative rationals, and the algorithmic memory is
represented exactly by
\begin{equation}\label{eq:joint-mass-representation}
 h_i^{(k)}=\sum_{s\in\mathcal R^{(k)}}
 A^{(s,k)}t^{(s)}\gamma_i^{(s)},\qquad
 H^{(k)}=\sum_i h_i^{(k)}.
\end{equation}
\item The quantitative memory bounds are
\begin{equation}\label{eq:joint-memory-bounds}
 h_i^{(k)}\ge0,\qquad 0\le H^{(k)}\le L_0,\qquad
 v_i^{(k)}=mh_i^{(k)}-H^{(k)}\le R_0H^{(k)}
 \quad(i\in[m]).
\end{equation}
\item Every $s\in\mathcal R^{(k)}$ satisfies the original-column
inequality
\begin{equation}\label{eq:joint-column-bounds}
 \sum_{i\ne j}\gamma_i^{(s)}d_{ij}^{(s)}(\vct y_j)
 \le\dfrac{L_{\rm col}}m
 \quad\text{for every }j\text{ and every }\vct y_j\in\Delta(A_j).
\end{equation}
The deficits $d_{ij}^{(s)}$ are evaluated at the \emph{original}
base profile and deviation of generation $s$, not at the current
product profile.
\item There are pairwise disjoint augmented certificate events
$E_{s,i}^{(k)}$ with the probabilities and conditional laws of
Lemma~\ref{lem:certificate-coupling}. In particular, their current
probabilities are $A^{(s,k)}\alpha_i^{(s)}$, their conditionally
sampled original deviations survive later updates, and the same
original law remains available for each opponent override.
\end{enumerate}
Every nonterminal iteration consequently has $\mu^{(k)}>5d$.
A successful LP preserves all four assertions, and a failed LP
executes a legal repair that preserves all four assertions while adding
one new certified generation. In particular, once $H^{(k)}>0$, all
later iterates have positive memory.
\end{lemma}

\begin{proof}
We prove the four assertions together, using the induction hypothesis
at iteration $k$ to verify the prerequisites of the local estimates
for that iteration.

\emph{The Base State.}
By Lemma~\ref{lem:init}, the initial strategy $\vct x^{(0)}$ is a
rational product profile. Algorithm~\ref{alg:main} sets
$\vct h^{(0)}=\vct0$ and $H^{(0)}=0$, while
$\mathcal R^{(0)}=\varnothing$. The sum in
\eqref{eq:joint-mass-representation} is empty; hence it is zero
coordinatewise, as required. We also have $v_i^{(0)}=0$ and thus
\eqref{eq:joint-memory-bounds}. Both the historical-column condition
and the existence of disjoint historical events are vacuous. This
establishes the base case.

\emph{The Descent Threshold Is Well Defined at the Induction State.}
Assume (I1)--(I4) at iteration $k$ and suppose the stopping test has
not passed. Thus $\varsigma=\sigma^{(k)}<\sigma$, and by (I2)
$0\le H\le L_0$. Bound~\eqref{eq:finite-threshold-bounds} gives $S_u'(H)>0$ on
this interval and $S_u(0)=G_0$. The uniform finite-step estimate
\eqref{eq:finite-strict-reserves} therefore yields
\begin{equation}\label{eq:joint-margin}
 \mu=S_u(H)-\varsigma-d
 \ge G_0-\sigma-d>5d>0.
\end{equation}
Hence the linearized LP is evaluated with a strictly positive,
fixed rational threshold, and its success and failure cases exhaust the
possibilities. If the stopping condition already holds, no additional
transition is required.

\emph{Case 1: The LP Succeeds.}
The primal LP~\eqref{eq:linearized-lp} supplies a nonnegative $\vct z$ with $Z\le1$.
Since $0<\eta<1$, all factors $1-\eta Z_j$ are positive and
\begin{equation}\label{eq:joint-descent-retention}
 1-\eta\le P=\prod_j(1-\eta Z_j)\le1.
\end{equation}
The profile update in~\eqref{eq:algorithm-descent-update} is
coordinatewise a convex combination of rational probability vectors,
so it preserves the rational product-profile property. By~\eqref{eq:algorithm-descent-memory},
$h_i^+=Ph_i$ and $H^+=PH$. Therefore
$$
 v_i^+=mh_i^+-H^+=P(mh_i-H)=Pv_i.
$$
Nonnegativity of the memory, its total cap, and its coordinate cap
follow immediately from $0<P\le1$:
$h_i^+\ge0$, $H^+\le L_0$, and
$v_i^+\le PR_0H=R_0H^+$.

For every previously stored generation, set
$A^{(s,k+1)}=PA^{(s,k)}$. Its contribution to the right-hand side
of~\eqref{eq:joint-mass-representation} becomes
$PA^{(s,k)}t^{(s)}\gamma_i^{(s)}$, so the sum is exactly
$Ph_i=h_i^+$. No new generation is inserted and no original
base profile or activation law is changed. Thus (I1) and (I3) hold at
iteration $k+1$.

For (I4), intersect each old event with the independent all-base
descent cylinder. That cylinder has probability $P$ and forces every
new coordinate to use its old strategy leaf. The intersection retains
the original conditional action law and multiplies every event mass by
$P$, as in Lemma~\ref{lem:certificate-coupling}. Intersecting disjoint
events with the same cylinder cannot create an overlap. The
opponent-override construction remains valid because the new all-base
labels are independent of the action leaf being replaced. This verifies
all four assertions after a successful LP.

\emph{Case 2: The LP Fails; Obtaining the Current Dual Certificate.}
By Lemma~\ref{lem:descent} and the positivity
in~\eqref{eq:joint-margin}, the failed LP has a rational optimal
dual distribution $\vct\lambda$ and deviations $(\vct b_i)_i$ for
which~\eqref{eq:dual-columns} and~\eqref{eq:dual-base} hold.
Define $\pi_i=m\lambda_i$ and $r=\max_i\pi_i-1$. Since the
$\lambda_i$ form a probability law, $r\ge0$. The diagonal part of
the dual certificate and nonnegative deficits give
$$
 \lambda_j\le F_j(\vct b_j)
 <\dfrac{\varsigma+\mu}{m}
 =\dfrac{S_u(H)-d}{m}<\dfrac3m.
$$
Thus $0\le r<2$. Furthermore~\eqref{eq:dual-columns} gives, for every
current column and every mixed override,
\begin{equation}\label{eq:joint-fresh-column}
 \sum_i\lambda_i d_{ij}(\vct y_j)
 <\dfrac{S_u(H)-d}{m}<\dfrac3m.
\end{equation}
\emph{Using Old Certificates to Establish the Floor Condition.}
By (I2) we have $mh_i\le(1+R_0)H$ and may put
$Y=R_0H$ in Lemma~\ref{lem:history-overlap}. Its hypotheses are
exactly the failed dual just obtained and the old certificates in
(I3)--(I4). The lemma produces the historical overlap inequality
\eqref{eq:historical-overlap}. If $H>0$, set $\psi=\tau/H$,
where $\tau=\sum_i\lambda_i v_i$; otherwise interpret $H\psi$ as
$\tau=0$. Because $h_i\ge0$, the mean-one variables
$\zeta_i=mh_i/H$ satisfy
$$
 \psi=\sum_i\lambda_i\zeta_i-1\le r
 \quad\text{when }H>0.
$$
For $H=0$ this inequality is not needed: the product $H\psi$
vanishes in every operational expression. Bound~\eqref{eq:historical-overlap} and the
value $L_{\rm col}=9$ give
$$
 0\le\vartheta\le\dfrac{(10+R_0)H}{m}
 <\dfrac{11}{m},\qquad
 (1+r)(1-\vartheta)+H+\tau<S_u(H)-d.
$$
Here $H\le L_0<1$ and $R_0<1$. Expanding the second inequality,
using $r<2$, and applying $m\ge66/d$ from~\eqref{eq:finite-player-cutoff} gives
$$
 r+\tau<C_u(H)-d+(1+r)\vartheta
 \le C_u(H)-d+\dfrac{33}{m}
 \le C_u(H)-\dfrac d2.
$$
For positive $H$, the bound
$$
 \dfrac{1+QH}{1+H}(r+\tau)-(r+Q\tau)
 =\dfrac{(Q-1)H}{1+H}(r-\psi)\ge0
$$
follows from the bound $\psi\le r$ just established and $Q>1$.
At $H=0$, we instead use $\tau=0$ directly, so the two expressions
$r+Q\tau$ and $r+\tau$ coincide; no value for $\psi$ is required.
Since $C_u(H)=R_0(1+H)/(1+QH)$, we conclude
\begin{equation}\label{eq:joint-floor}
 r+Q\tau\le R_0.
\end{equation}
The old-generation certificates thus establish the new floor condition.

\emph{Legal Tilt and the Fresh Stored Column.}
Apply Lemma~\ref{lem:finite-legal-tilt} using (I2) and
\eqref{eq:joint-floor}. It shows that
$\vct\gamma$ is a rational probability law, that
$0\le\gamma_i<3/m$ and $\gamma_i\le3\lambda_i$, and that
the reverse allocations $\delta_i$ are nonnegative and sum to
$1-\rho$. The denominator $Z_0$ is positive: indeed
$Z_0=G_0-Q\tau\ge1+r\ge1$. Also
$G_0-Qv_i>1$ because $v_i\le R_0L_0$ and
$QL_0<c<1$. Thus every probability used in the activation is
well defined; $t\gamma_i<1$ follows from the cutoff~\eqref{eq:finite-player-cutoff}.

We can now establish the fresh column bound. Since all deficits are
nonnegative, $\gamma_i\le3\lambda_i$ and
\eqref{eq:joint-fresh-column} imply
\begin{equation}\label{eq:joint-new-certificate}
 \sum_{i\ne j}\gamma_i d_{ij}(\vct y_j)
 \le3\sum_i\lambda_i d_{ij}(\vct y_j)
 <\dfrac9m=\dfrac{L_{\rm col}}m
\end{equation}
for every $j$ and every $\vct y_j\in\Delta(A_j)$.
The bound uses the original current base profile $\vct x^{(k)}$,
which becomes the recorded base profile of the new generation.
The factor three applies to the fresh comparison~\eqref{eq:joint-new-certificate}.

\emph{The Two Sweeps, Correction, and Certificate Retention.}
The nonnegative reverse allocations sum to $1-\rho<1$.
Hence every prefix $P_{i-1}=1-\sum_{j<i}\delta_j$ is at least
$\rho>0$, and $\delta_i\le P_{i-1}$, so the reverse
best-response sweep~\eqref{eq:algorithm-reverse-sweep} defines valid rational
product mixtures. The raw profile has nonnegative fixed correction deficits
defined in~\eqref{eq:algorithm-correction}; these are frozen before the forward
sweep~\eqref{eq:algorithm-forward-correction}. The hypotheses of
Lemma~\ref{lem:finite-raw-error} are now satisfied: (I3)--(I4)
provide the old columns and their event laws, while~\eqref{eq:joint-new-certificate}
provides the new column and Lemma~\ref{lem:finite-legal-tilt} the legal tilt. Lemma~\ref{lem:finite-raw-error} yields
$$
 0\le E\le\dfrac{120u}{m},\qquad
 \widehat C=\dfrac{C_{\rm raw}}{1+E}\ge c,\qquad
 \varsigma^+\ge c\varsigma+D-E.
$$
All forward correction probabilities in~\eqref{eq:algorithm-forward-correction} have the form
$e_i/(1+\sum_{j\le i}e_j)$ and therefore belong to $[0,1)$.
The new product profile is rational, since the correction deficits
are rational payoff expressions and the best responses are pure.

By Lemma~\ref{lem:certificate-coupling}, we may intersect old events
with the current all-base and correction-base cylinders. The resulting
common mass is $\widehat C$ times the preceding mass. Thinning using an
independent auxiliary certificate coin with success probability
$c/\widehat C$ changes that factor to exactly $c$ but does not change
any output strategy or conditional payoff law. Consequently
\begin{equation}\label{eq:joint-new-masses}
 A^{(s,k+1)}=cA^{(s,k)}\quad(s\in\mathcal R^{(k)}),
 \qquad A^{(k,k+1)}=c.
\end{equation}
The newly constructed events are disjoint from each other and from
all retained old events: an old event requires the current generation's
base labels, while a new activation requires an activation label at its
own coordinate and base labels at the opponent coordinates. The new generation
satisfies~\eqref{eq:joint-new-certificate}; for every old generation,
its historical column is unchanged because both its original base
profile and its original mixed deviation are frozen. This proves
(I3)--(I4) at the next iteration.

Combining~\eqref{eq:joint-mass-representation} and~\eqref{eq:joint-new-masses}
gives the exact mass formula, including the just-created generation:
\begin{equation}\label{eq:joint-repair-mass}
 h_i^+=\sum_{s\in\mathcal R^{(k)}}
 cA^{(s,k)}t^{(s)}\gamma_i^{(s)}+ct\gamma_i
 =c(h_i+t\gamma_i),\qquad H^+=c(H+t).
\end{equation}
This is the update~\eqref{eq:algorithm-repair-memory} performed by the algorithm, so the
auxiliary certificates agree with its memory variables. Nonnegativity of
$h_i^+$ follows from $h_i,\gamma_i\ge0$, and
$$
 H^+=c(H+t)\le c(L_0+t)=L_0,
$$
where the last equality uses the definition $L_0=ct/(1-c)$
in~\eqref{eq:finite-potential-controls}.

It remains to verify the centered bound in~\eqref{eq:joint-memory-bounds}.
Write $\chi_i=m\gamma_i$ and $Q_0=tQ$. The legal tilt~\eqref{eq:finite-tilt} gives
$\chi_i\le G_0-Qv_i=1+R_0-Qv_i$. Therefore
$$
 v_i^+=c[v_i+t(\chi_i-1)]
 \le c[(1-Q_0)v_i+tR_0]
 \le cR_0[(1-Q_0)H+t]
 \le R_0H^+.
$$
The penultimate inequality uses $1-Q_0>0$ and the induction
hypothesis $v_i\le R_0H$; the last uses $Q_0>0$ and
\eqref{eq:joint-repair-mass}. Together with
the nonnegativity and total-memory bound established above, this proves (I1)--(I2) for the repair
transition. Every assertion at iteration $k+1$ is therefore obtained
from the induction hypothesis at iteration $k$ and a local estimate
whose premises have already been checked.

\emph{Persistence of Positive Memory.}
A descent multiplies $H$ by $P\ge1-\eta>0$ according
to~\eqref{eq:joint-descent-retention}; a repair replaces it by $c(H+t)>0$
according to~\eqref{eq:joint-repair-mass}, since $c,t>0$. Thus positive memory remains positive
thereafter. Starting from zero memory, every successful descent leaves
it at zero, and the first repair creates $H^+=ct>0$. This proves the
final assertion and completes the simultaneous induction.
\end{proof}

\paragraph{Auxiliary Certificates and Implementation.}
The coupling of Lemma~\ref{lem:certificate-coupling} uses an augmented
probability space to certify inequalities. Its common masses,
original-column certificates, and stored generations are witnesses for
the analysis. At each iteration the algorithm stores a rational product
profile and the $m$ rational memory weights $h_i$, so the running-time
proof in Subsection~\ref{subsec:runtime} counts these rational data.

\subsection{Iteration Bound and Bit Complexity}\label{subsec:runtime}

Finally, we turn the preceding progress estimates into a bound on the
running time. We combine Proposition~\ref{prop:potential} with the
reachable-state guarantees of Lemma~\ref{lem:joint-invariant} to bound
the number of iterations, then analyze the rational arithmetic cost
to complete the proof of Proposition~\ref{prop:runtime}.

\begin{proof}[Proof of Proposition~\ref{prop:runtime}]
Let \(\Delta_0=4\eta d\).  While \(H=0\), one has
$S_u(H)=G_0$ and $\mu=G_0-\varsigma-d$, hence
$\varsigma+\mu=G_0-d<2$. Therefore the descent bound~\eqref{eq:descent-bound} gives
\[
 \varsigma^+-\varsigma
 \ge\eta\mu-\eta^2(\varsigma+\mu)
 \ge5\eta d-2\eta^2\ge\Delta_0 ,
\]
because $G_0<2$ by~\eqref{eq:finite-parameter-approximation}, $\mu>5d$
by~\eqref{eq:joint-margin}, and $\eta\le d/24$
by~\eqref{eq:algorithm-primary-controls}; a failed step is
the first repair. Since \(\varsigma^{(0)}\ge1\), this phase has at most
\(1+(\sigma-1)/\Delta_0\) iterations.

The zero-memory phase is unique:
Lemma~\ref{lem:joint-invariant} gives $H^+=PH=0$ for a passing step
at $H=0$, whereas the first failed step gives $H^+=ct>0$.
All later steps preserve positive memory, and every passing step in
the zero-memory phase consumes at
least $\Delta_0$ of the remaining gap to the fixed target $\sigma$.
This phase ends upon reaching $\sigma$ or executing the first repair.

After the first repair, $H=ct>0$ by~\eqref{eq:joint-repair-mass}.
For the potential~\eqref{eq:finite-tilted-potential} at that state, the bounds
\(\varsigma\ge0\), \(0<a_u<1\), and \(cu<1\) give
\[
 L_{\min}:=-\dfrac{G_0}{ct}+\log(cu)
 \le\widehat\Phi_u.
\]
Proposition~\ref{prop:potential} shows that every subsequent positive-memory
iteration strictly increases $\widehat\Phi_u$. Hence this lower bound
propagates to all later states.  For every nonterminal state, the
$1/H$ term is negative because $G_0>\sigma+6d$
by~\eqref{eq:finite-strict-reserves}, and
\(\log(H/f_0)<0\); consequently
\[
 \widehat\Phi_u\le a_u\log(1+QL_0)<\log2<1.
\]
The potential interval therefore has length at most the fixed constant
\begin{equation}\label{eq:runtime-potential-range}
 R_\Phi:=1-L_{\min}=1+\dfrac{G_0}{ct}-\log(cu).
\end{equation}
For the lower bound, recall that $H=ct$ immediately after the first
repair. At that state, the reciprocal contribution
$(\varsigma-G_0+2d)/H$ is at least $-G_0/(ct)$ because
$\varsigma\ge0$; the term $(1-a_u)\log(H/f_0)$ is at least
$\log(cu)$ since $0<1-a_u<1$ and $\log(cu)<0$; and the last
logarithmic term is nonnegative.

The upper bound is uniform over \emph{all} positive-memory
nonterminal states, including those immediately following a repair.
Indeed $\varsigma<\sigma$ and $G_0>\sigma+6d$ imply
$\varsigma-G_0+2d<-4d$, and so the reciprocal term is strictly
negative. Also $H\le L_0<f_0$ makes $\log(H/f_0)<0$, while
$QL_0<c<1$ gives $\log(1+QL_0)<\log2$.
Thus the potential range depends only on the fixed target $\sigma$.
The logarithms enter only this real-valued termination argument.

Proposition~\ref{prop:potential} and the memory cap~\eqref{eq:joint-memory-bounds}
show that every positive-memory iteration increases the potential by at least
$$
 \kappa:=\min\left\{\dfrac{\eta d}{2L_0},
                    \dfrac{3d(1-c)}{L_0}\right\}>0 .
$$
There are at most \(\lceil R_\Phi/\kappa\rceil+1\) such iterations.
The finite inequalities of Proposition~\ref{prop:potential} apply to each
step: every successful step raises the potential by at
least $\eta d/(2H)$, and every actual repair raises it by at least
$3d(1-c)/H$. The inductive bound $H\le L_0$ converts both into
the common positive increment $\kappa$. The strict bounds rule out
an infinite run even if the LP alternates arbitrarily between the
two branches.
Together with the $H=0$ bound, this gives the constant
\begin{equation}\label{eq:runtime-iteration-bound}
N_0:=1+\lceil(\sigma-1)/\Delta_0\rceil
 +\lceil R_\Phi/\kappa\rceil+1.
\end{equation}
It also bounds the number of
stored repair generations.  This constant depends only on $\sigma$,
because all controls were fixed before running the algorithm.

It remains to verify the bit bound.
The controls $d,f_0,u,\eta,m_0$, together with
$c,\rho,t,\beta,\nu$ and $R_0,G_0,Q,a_u,L_0$,
are chosen effectively from the fixed rational target $\sigma$ before
the game is processed. The rational search in
Subsection~\ref{subsec:proof-controls} terminates, and its result
may be fixed once for that target. Therefore the bit lengths of these
parameters are constants depending on $\sigma$, not on the input
length $B$. This observation also makes the small-player cutoff
branch of Algorithm~\ref{alg:main} effective.

The LP~\eqref{eq:linearized-lp} in one round has at most
\(\sum_j|A_j|+1\le B+1\) variables and at most
\(\sum_i|A_i|+1\le B+1\) constraints.  If all current rational
coefficients have bit length at most \(L\), every coefficient formed by
a contraction, a best-response comparison, or a fixed correction has bit length bounded by a fixed polynomial in \(B+L\).
A rational LP has an optimal basic feasible solution; by the determinant
bound (Hadamard's inequality, or Cramer's rule), when its coefficients
have bit length at most \(L'\) and the LP has dimension at most \(B+1\), the
coordinates of such a solution and of an optimal dual have bit length
at most \(O\left((B+1)^2\bigl(L'+\log(B+1)\bigr)\right).\)

More explicitly, an expected payoff is a sum of rational products,
with at most one term per explicitly listed pure-profile entry; each
product contains at most $m$ strategy probabilities and one input
payoff. Since $m$ and the number of listed entries are at most
polynomial in $B$, contractions and deviation comparisons produce
rationals whose numerator and denominator lengths are polynomial in
$B+L$. The activation weights~\eqref{eq:repair-controls},
reverse-budget prefixes~\eqref{eq:algorithm-reverse-sweep}, and
frozen correction deficits~\eqref{eq:algorithm-correction} are obtained by polynomially many rational
additions, multiplications, and divisions. Their denominators never
vanish, by Lemma~\ref{lem:finite-legal-tilt} and the correction
rule~\eqref{eq:algorithm-forward-correction}: $Z_0\ge1$, $P_{i-1}\ge\rho>0$,
$1+\sum_{j\le i}e_j\ge1$, and $1-t\gamma_i>0$.
The independent certificate coins and conditional label trees in
Lemma~\ref{lem:certificate-coupling} are used only in the proof.

Thus, if \(L_k\) denotes the
largest state bit length after round \(k\), then
\(
L_{k+1}\le p(B+L_k),
\) where $p$ is a fixed polynomial with nonnegative coefficients, depending only on the fixed target $\sigma$, and independent of $k$ and the input game.
Since $N=N(\sigma)$ is the fixed bound~\eqref{eq:runtime-iteration-bound}, induction gives
\(L_k\le B^{O_\sigma(1)}\) for \(k\le N\).  Each tensor contraction and
best-response scan is polynomial in the explicit input size and these
bit lengths, and each sweep has \(O(\sum_i|A_i|)\) steps.  Therefore all
rounds together take \(B^{O_\sigma(1)}\) time.
\end{proof}

\section{Conclusion}
\label{sec:conclusion}
In this paper, we developed a polynomial-time algorithm
for computing approximate Nash equilibria in general multiplayer
games, achieving a $(1-\sigma/m)$-approximation for every fixed
$1<\sigma<\sigma_\star$, where $\sigma_\star\approx1.5867$,
and all sufficiently large $m$. This result breaks the long-standing
one-player extension barrier by improving the leading coefficient
of the classical $1-1/m$ guarantee. Our main contribution is an
iterative descent-and-repair framework that combines dual
certificates with historical information to enable repeated repairs,
while the potential function, motivated by
continuous-time dynamics, guarantees global progress and polynomial-time
termination. 
Our analysis suggests that exploiting historical information across
iterations provides a promising direction for further improving
approximation guarantees in multiplayer games.

\section*{AI Methodology}
We used OpenAI's ChatGPT to assist with refining the probability distribution used in the repair procedure and computing the optimal coefficient. The authors take full responsibility for the correctness and originality of the results presented in this paper.

\bibliographystyle{alpha}
\bibliography{main}

\end{document}